\documentclass{mathincs}
\usepackage[utf8]{inputenc}
\usepackage[T1]{fontenc}
\usepackage{amssymb,amsmath,xspace,hyperref,graphicx,multirow}
\usepackage{epsfig,setspace,acronym,array,graphicx}
 \usepackage{url}
 \usepackage{mdframed}
\usepackage{listings}

 \newtheorem{thm}{Theorem}[section]
   \newtheorem{goal}{Goal}[section]
  \newtheorem{theorem}{Theorem}[section]

 \theoremstyle{definition}
 
  \newtheorem{definition}[thm]{Definition}
  \newtheorem{constraint}{Constraint}[section]
 \theoremstyle{remark}

 \numberwithin{equation}{section}

\title{Formal Analysis of Optical Systems}
\author[S.Kh.Afshar, U.Siddique, M.Y.Mahmoud, V.Aravantinos, O.Seddiki, O.Hasan and S.Tahar]{Sanaz Khan-Afshar \and Umair Siddique \and Mohamed Yousri Mahmoud \and Vincent Aravantinos \and Ons Seddiki \and Osman Hasan \and Sofi\`{e}ne Tahar}
\address {
  Department of Electrical and Computer Engineering \\
  Concordia University\ \\
  Montr\'{e}al, Qu\'{e}bec, Canada
}
\email{\\ \{s\_khanaf,muh\_sidd,mo\_solim,vincent,o\_sed,o\_hasan,tahar\}@ece.concordia.ca}

\def\hol#1{$\mathtt{#1}$}

\def\Leftrightarrowdef{\Leftrightarrow}

\def\emf{emf}
\let\vdot=\cdot
\def\Lambda#1{\lambda #1\qdot}
\def\wofwave{\omega\_of\_w\ }
\def\eofwave{e\_of\_w\ }
\def\Exists#1{\exists #1\qdot}
\def\mapt{map\_triple\ }
\def\hofwave{h\_of\_w\ }
\def\planewave{plane\_wave\ }
\def\eqdef{=}
\def\NS{\negthinspace\negthinspace}
\def\isplanewaveatitf{is\_plane\_wave\_at\_interface\ }
\def\isplanewave{is\_plane\_wave\ }
\def\isvaliditf{is\_valid\_interface\ }
\def\isinplan{is\_in\_plane\ }
\def\kofwave{k\_of\_w\ }
\def\boundaryconditions{boundary\_conditions\ }

\def\vdashdef{\vdash}
\def\qdot{.\ }
\def\Forall#1{\forall #1\qdot}

\def\aresymetric{are\_sym\_wrt\ }

\def\boundaryconds{boundary\_conditions\ }
\def\eofemf{e\_of\_emf\ }
\def\hofemf{h\_of\_emf\ }

\def\Q#1#2{\frac{\partial #1}{\partial #2}}
\def\QQ#1#2{\frac{\partial^2 #1}{\partial #2^2}}

\def\lrarrow{~\lower.2ex\hbox{$\rightarrow$}\kern-2.4ex\raise.7ex\hbox{$\leftarrow$}~}
\def\rlarrow{~\lower.2ex\hbox{$\leftarrow$}\kern-2.3ex\raise.7ex\hbox{$\rightarrow$}~}

\def\labelenumii{\Roman{enumii}.}
\def\theenumii{\Roman{enumii}}
\def\p@enumii{\theenumi}
\def\tt#1{\texttt{#1}}

\def\hol#1{$\mathtt{#1}$}
\def\vectt#1{$\vec{\mathtt #1}$}

\def\shol#1{\hol{{\ #1}}}
\def\Leftrightarrowdef{\ \Leftrightarrow\ }

\def\emf{emf\ }
\let\vdot=\cdot
\def\Lambda#1{\lambda #1\qdot\ }
\def\wofwave{\omega\_of\_w\ }
\def\eofwave{e\_of\_w\ }
\def\Exists#1{\exists #1\qdot\ }
\def\mapt{map\_trpl\ }
\def\hofwave{h\_of\_w\ }
\def\planewave{plane\_wave\ }
\def\eqdef{\ =\ }
\def\NS{\negthinspace\negthinspace}
\def\isplanewaveatitf{is\_plane\_wave\_at\_int\ }
\def\isplanewave{is\_plane\_wave\ }
\def\isvaliditf{is\_valid\_interface\ }
\def\isinplan{is\_in\_plane\ }
\def\kofwave{k\_of\_w\ }
\def\boundaryconditions{boundary\_conditions\ }

\def\normalofint{normal\_of\_interface\ }

\def\vdashdef{\vdash_{def} \ }
\def\qdot{.\ }
\def\Forall#1{\forall #1\qdot\ }

\def\aresymetric{are\_sym\_wrt\ }

\def\boundaryconds{boundary\_conditions\ }
\def\eofemf{e\_of\_emf\ }
\def\hofemf{h\_of\_emf\ }
\def\nonnull{non\_null\ }

\def\real{\mathbb{R}}
\def\complex{\mathbb{C}}
\def\labelenumii{\Roman{enumii}.}
\def\theenumii{\Roman{enumii}}
\def\p@enumii{\theenumi}

-lmtk10
\font\bitt=rm-lmtko10



\begin{document}

  \begin{abstract}

Optical systems are becoming increasingly important by resolving many
bottlenecks in today's communication, electronics, and biomedical systems.
However, given the continuous nature of optics, the inability to efficiently analyze optical system models using traditional paper-and-pencil
and computer simulation approaches sets limits especially in safety-critical
applications. In order to overcome these limitations, we propose to employ
higher-order-logic theorem proving as a complement to computational and numerical
approaches to improve optical model analysis in a comprehensive
framework.
The proposed framework allows formal analysis of optical systems at four abstraction levels, i.e., ray, wave, electromagnetic, and quantum.

  \end{abstract}

 \maketitle

  \section{Introduction} \label{sec_intro}

  Thanks mainly to its ability to provide high capacity communication links, optical technology is increasingly being exploited in applications ranging from ubiquitous Internet and mobile communications to more advanced scientific domains, such as programmable integrated platforms where processors are connected through optical networks, bio-photonics and laser material processing. The accuracy of operation for such optical systems is very important due to the financial and/or safety critical nature of their applications. Optical technology also has unique properties that make it extremely useful in medicine: laser surgeries are replacing traditional scalpel based surgeries for removing tumours, curing deafness and spine injuries. The minor bugs in optical systems
 can, however, lead to disastrous consequences such as the loss of human lives because of their use in surgeries and high precision biomedical devices, or financial loss because of their use in high budget space missions. For example, the Hubble Telescope \cite{Hubble_90}, which is considered as one of NASA's  largest projects with a budget of \$1.6 billion, faced  a historical system failure due to the misalignment of  two mirrors of the telescope. In practice, a significant portion of the design time is spent on analyzing every aspect of the design, so that functional errors can be caught prior to the production of the actual device.

The verification of an optical system is generally achieved by combining various means. The most basic one is the actual manufacturing of a prototype that can then be tested. However, this is obviously a costly technique. Therefore, engineers try as much as they can to detect faults in a design before resorting to testing. This requires developing a mathematical model of the system and then analyzing it.
Such a model is
based on various theories of physics depending upon the system properties that need to be verified. The simplest theory is ray optics \cite{Fund_of_Photonics}, which considers light as a simple geometric line whose orientation changes according to the medium changes. Wave optics \cite{Fund_of_Photonics}, which considers light as a scalar wave, allows a more detailed analysis of optical systems. This allows taking into account phenomena like diffraction. A more enhanced theory is
electromagnetic optics \cite{EM_REF}, which models light as an electromagnetic wave driven by Maxwell equations, addressing phenomena like polarization and dispersion of light. Finally, the theory of quantum optics \cite{Quantum_ref} considers light as a stream of photons, whose behaviour is driven by the laws of quantum mechanics. The choice of theory primarily depends on the system and the specifications that we want to verify: for instance, checking that a given optical resonator is stable can be achieved very simply and reliably with ray optics, however, ensuring that no energy is lost when light travels through a waveguide requires electromagnetics. On the other hand, modelling photonic devices (e.g., a laser or light detector) requires quantum optics.

In general, the analysis of optical systems is carried out using three techniques: paper-and-pencil based proofs, computer simulations, and computer algebra systems. In the \textit{paper-and-pencil proof}, a mathematical model of the  optical system is built using the underlying physical concepts. This model is then used to verify that the system exhibits the desired properties using mathematical reasoning on paper \cite{Lasers_66,Physica_review_ABCD,TimeDomain_ABCS_1998}.
However, considering the complexity of present-age optical and laser systems, such an analysis is very difficult if not impossible, and thus quite error-prone. Many examples of erroneous paper-and-pencil proofs are available in the open literature, a recent one can be found in \cite{naqvi_corrected_10} and its identification and correction is reported in \cite{A_Naqvi_11}.

The main idea of \textit{simulation-based methods} is to construct a discretized model and then simulate the output of the system using different input patterns.
 In ray optics, one of the most commonly used computer-based analysis techniques is the numerical computation of complex ray-transfer matrices \cite{LASCAD,rezonator,Numerical_ABCD_2011}. In electromagnetic optics, we can refer to many works on computational methods in electromagnetism \cite{garg_08}, e.g. \cite{Johnson_01} and \cite{Liu_06}. In case of quantum optics, the simulation based analysis cannot be performed by ordinary computers \cite{Simulatingphysics}. However, some tools for quantum systems analysis have been developed based on numerical computations, e.g., \cite{matlabtoolbox}. One of the disadvantages of computer simulations is the tremendous
amount of CPU time and memory that are generally required to
reach usable meaningful results \cite{Heinbockel_04}. In \cite{Hayes_99,Johnson_01,Liu_06,Yin_99},
the authors discuss different methodologies to improve the memory consumption and speed
of numerical approaches; however, computer simulation techniques still fail to provide perfectly accurate results due to the heuristics and approximations of the underlying numerical algorithms.

Finally, \textit{computer algebra systems (CAS)} \cite{OPTICA} are becoming quite popular for the analysis of optical systems. In CASs, mathematical computations are done using symbolic
algorithms which are better than simulation-based analysis in terms of precision. But the simplification performed by computer algebra systems are not 100 \% reliable \cite{harrison_thesis} due to their inability to deal with side conditions, which are necessary for a mathematical expression to be valid, e.g., $x \neq 0$ is a side condition for expression $\frac{x}{x}$. Another source of inaccuracy in computer algebra systems is the presence of the unverified huge symbolic manipulation algorithms in their core, which are quite likely to contain bugs.

As a solution to enhance the accuracy of system analysis, we propose to use \emph{formal methods},
besides traditional approaches, as a complementary technique. The main idea behind formal methods is to develop a mathematical model for the given system and analyze this model using computer-based mathematical reasoning, which increases the chances of catching subtle but critical design errors that are often ignored by traditional techniques. There are essentially two  main formal verification techniques: \textit{model checking} \cite{modelchecking_book} and \textit{theorem proving} \cite{harrison_book}. Model checking is an automated verification technique for  systems that can be expressed as finite-state machines. On the other hand, theorem proving is  generally an interactive verification technique, but it is more flexible and can handle a variety of systems. The continuous nature of optical systems prevents the  model from being abstracted within a finite-state machine without losing accuracy. Therefore, model checking cannot guarantee absolute correctness of analysis in the case of optical systems. On the other hand, theorem proving, based on higher-order logic, does not impose any expressiveness restriction and allows us to formalize optical system analysis fundamentals, like complex numbers, differentiation, transcendental functions, vector space analysis and Euclidean geometry. Therefore, the proposed framework for the formal optical system analysis is based on higher-order-logic theorem proving.

The rest of the paper is organized as follows: Section \ref{hol} briefly describes logic and theorem proving to facilitate the understanding of this work for the optical system analysis community. The proposed approach is illustrated in Section \ref{sec_proposal}. We provide some technical insights in using the proposed approach for analyzing optical systems at the ray, electromagnetic and quantum levels in Sections 4-6, respectively. In Section \ref{sec_app}, to demonstrate the effectiveness of employing formal methods to enhance the reliability of optical system analysis, we present formalization of stability analysis of two-mirror Fabry-P\'{e}rot resonators, based on ray optics. In Section \ref{sec_bridge}, we presented our preliminary results on how to connect a theorem prover (i.e., HOL Light in our work) to other mechanized mathematical systems to broaden the range of applications which can be addressed by our framework.
 In Section \ref{sec_CAS}, we highlight the engineering prospects of our research and provide an assessment of necessary steps required to build an infrastructure that is feasible to be used by the optics industry.
Finally, Section \ref{sec_disc} concludes the paper with a discussion on challenges perspectives we faced in the formalization of optical systems and some potential future directions.

\section{Higher-Order Logic and Theorem Proving}
\label{hol}
In general, a \emph{logic} provides a (formal) language to express mathematical facts, and a definition of what is a \emph{true sentence} in this language. For example, the most basic kind of logic is the \emph{propositional logic} (also called \emph{boolean logic}), which only allows sentences formed by propositional variables and boolean connectives:  \emph{and} (\hol{\wedge}), \emph{or} (\hol{\vee}), \emph{not} (\hol{\neg}), \emph{implies} (\hol{\Rightarrow}) and equality (=) connectives. For instance,  \hol{(A\Rightarrow B) \wedge (B\Rightarrow C) \Rightarrow (A\Rightarrow C)} is a sentence of propositional logic. In addition, one can easily see that it is a \emph{true} sentence (using the transitivity of implication).

Only the overall structure of mathematical sentences can be expressed in propositional logic and one lacks the ability to talk about objects and their properties. This problem is answered by first-order logic that introduces \emph{terms} (which formalize the notion of ``object'') and \emph{predicates} (which formalize the notion of ``property of an object''). Terms are built inductively from constants and functions, e.g., the set of natural numbers is built from the constant \hol 0\ and the function \hol{SUC}, hence, \hol 1\ is represented by \hol{SUC(0)}, \hol 2\ by \hol{SUC(SUC(0))}, etc. Being an even or a prime number are then properties of natural numbers that can be represented by predicates. First-order logic thus allows to write sentences like \hol{Even(0)} or \hol{Prime(SUC(SUC(0))}. In order to get even closer to the usual mathematical language, first-order logic also introduces the notion of a \emph{variable}, which allows for instance to write a sentence like: \hol{Even(x) \Rightarrow Even(SUC(SUC(x))}, where \hol x\ is a variable that can be replaced by any term representing a number. Finally, sentences with variables are not complete if we cannot specify how variables should be interpreted, so two new ways of building a sentence are added to the language by using \emph{for all} (\hol \forall ) and \emph{there exists}  (\hol \exists) (called \emph{quantifiers}): e.g., ``\hol{ \forall x.\ Even(x)\ \Rightarrow\ Even(SUC(SUC(x))}'' or ``\hol{ \exists x.\ Prime(x)}''.

First-order logic does not permit quantifying over predicates. For instance, it is impossible to express the induction principle for natural numbers:
\hol{\forall P. P(0)\wedge} \hol{ (\forall n. P(n) \Rightarrow P(SUC(n)))} \hol{ \Rightarrow \forall n. P(n)} since \hol \forall\ can only be applied to variables and not to predicates.
Higher-order logic provides this feature and thus, in comparison to the aforementioned logic is stronger to represent mathematical theories.

Given a logic, the most frequent problem is to try to determine whether a given sentence is true or not. This is done by considering a set of \emph{axioms}, i.e., basic sentences that are assumed to be true (e.g., \hol{P\vee\neg P}), and \emph{inference rules}, i.e., rules that allow to derive the truth of a sentence depending upon the truth of other sentences (e.g., if \hol P\ and \hol Q\ are true sentences, then \hol{P\wedge Q} is a true sentence). Using axioms and inference rules, one can thus \emph{prove} or disprove logical sentences. This idea is at the principle core of theorem proving: the language definition, the axioms and inference rules can be implemented in the theorem prover. This allows the user to write down mathematical sentences inside the theorem prover, and then to prove them \emph{using only} the axioms and inference rules provided by the theorem prover. This latter point is essential since,
assuming there exists no inconsistencies in the foundations of the theorem prover,
it ensures that no unsound reasoning step can be used to prove a theorem. This guarantees that any sentence, which is proved in a theorem prover, is indeed true.

\section{Proposed Approach}\label{sec_proposal}

\begin{figure}[h]
\centering
\includegraphics[width=5.5in]{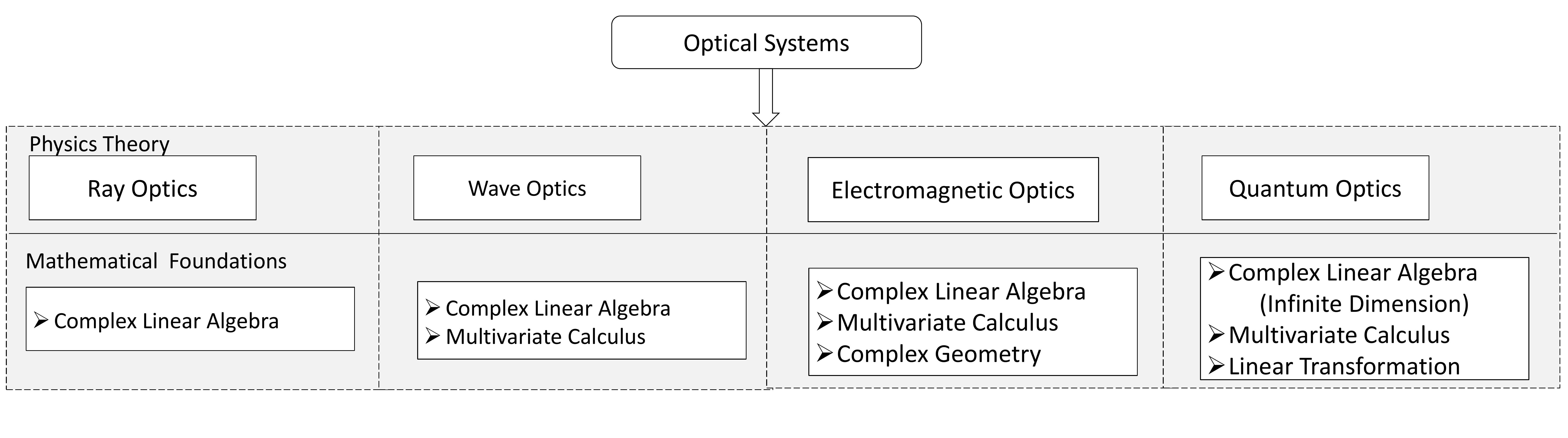} \vspace{-.1cm}
\caption{Mathematical Requirements for Optical System Analysis}
\label{fig:req} % labeling to refer it inside the text
\end{figure}

As described in Section 1, optical systems can be described by four theories, namely ray optics, wave optics, electromagnetic optics, and quantum optics. In this section, we focus on identifying the required mathematical foundations for the formal reasoning about these four theories. As depicted in Figure \ref{fig:req}, all these four theories of optics require complex linear algebra, and as we move from left to right in, the complexity of mathematical foundations increases because of the sophistication of underlying physics theories. Besides linear algebra, wave optics requires multivariate calculus, electromagnetic theories require support of complex geometry, and quantum optics involves infinite dimensional linear algebra and linear transformation.
Besides the identification of these foundational mathematical theories, we also have to choose a suitable theorem prover. This choice is primarily based on the available formalization related to the above mentioned theories to facilitate building upon the existing work instead of starting from scratch.
 Briefly, we had the option to choose between two proof styles of interactive theorem proving: the procedural style, where proofs are scripts of commands like in HOL Light, and the declarative style, where proofs are texts in a controlled natural language, like in Mizar. While in declarative style, formal proofs can be written similar to normal mathematical text, in procedural style, in general, it is easier to introduce automation \cite{freek_12}. One of our major goal in this project is making it usable in the field of optics, hence we preferred to introduce automation to our final product; we chose procedural style over declarative style of interactive theorem proving.
It is also worth mentioning that for most of procedural style theorem provers, a declarative mode has been developed, e.g., Isar mode of Isabell, Mizar mode of HOL, and Mizar Light in HOL Light.
Thus, we had three major reasons to choose HOL Light among other theorem provers like Coq, Isabell/HOL, and PVS; given the fact that neither of them had any development of complex vector analysis:

\begin{enumerate}
\item Rich libraries of complex analysis and vector analysis; which we extensively use to develop formal analysis of complex vectors,
\item Active projects like flyspeck \cite{flyspeck}; which constantly enrich the libraries on geometry,
\item The fact that vector analysis has been transferred from HOL Light to other theorem provers, e.g., Isabell/HOL \cite{holzl_13} was an assurance for us in case we needed to transfer our formalization to another platform.
 \end{enumerate}

 In \cite{hasan_09}, we presented the formal analysis of optical waveguide using the HOL4 theorem prover \cite{gordon_93}. This work was rather a feasibility study of applying theorem proving in the domain of optics with the specific example of rectangular waveguides. Moreover, it was primarily based on the real
analysis which is insufficient to capture the dynamics of most optical and photonic systems. After \cite{hasan_09}, we realized that the recent developments of multivariate analysis formalization in the HOL Light  \cite{harrison-hollight} theorem prover is a far better choice for optical system analysis.
 Currently available multivariate analysis theories can be built upon to develop all the mathematical requirements, given in Figure \ref{fig:req}.

%\subsection{Main Objectives}

\begin{figure}[h]
\centering
\includegraphics[width=5in]{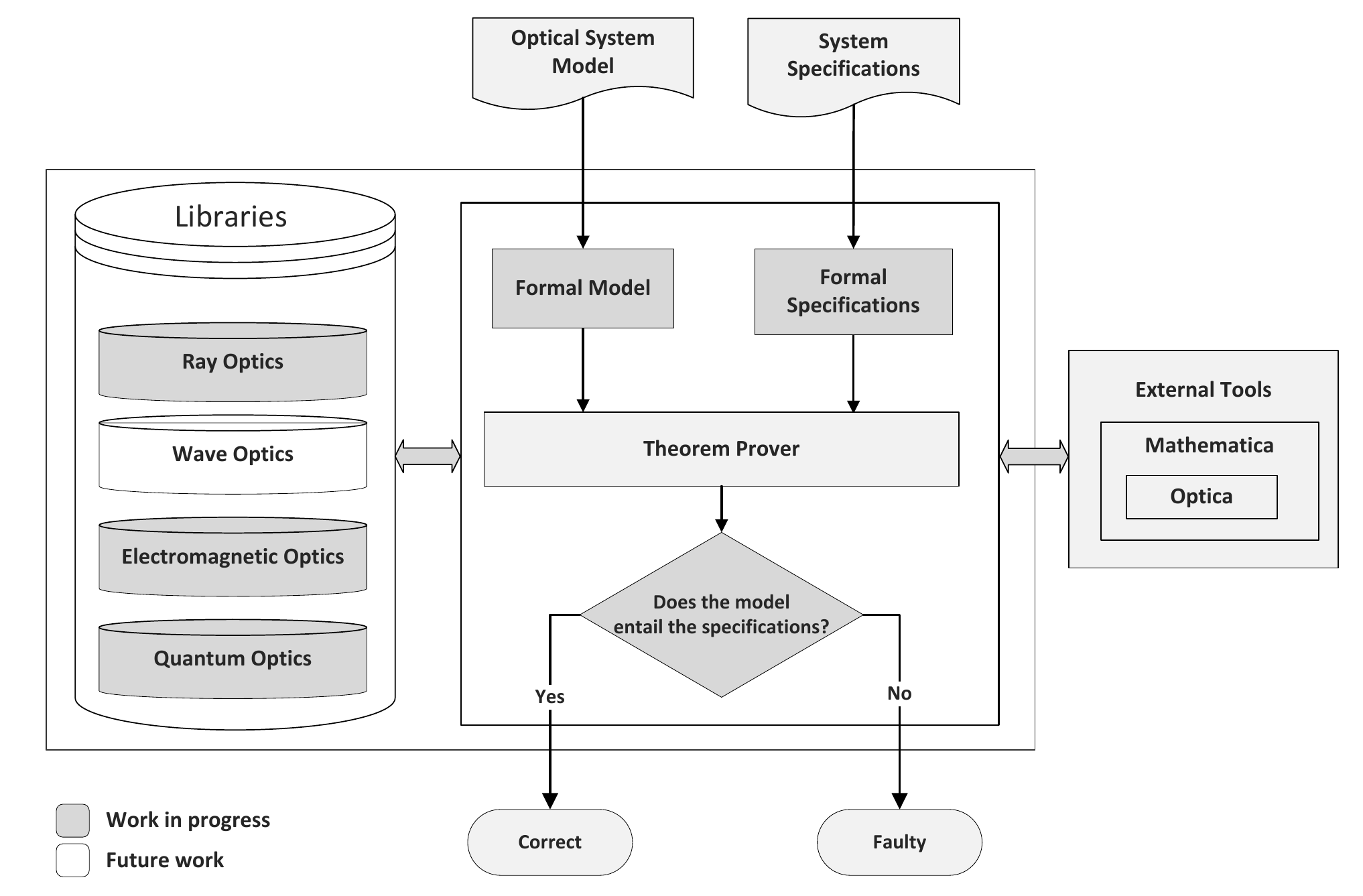}\vspace{-.1cm}
\caption{Proposed Formal Analysis Approach}
\label{fig:PF} % labeling to refer it inside the text
\end{figure}

Figure \ref{fig:PF} shows the major steps that should be taken to
formally verify an optical system. In general, before verifying any
system in a theorem prover, two sets of operations should be
completed. These two are referred to as  \emph{formal specification}
and \emph{formal modelling} of the system.
Once both the specification of the system (in terms of properties), and implementation of the system are formally described, the system can be termed as completely modelled in higher-order-logic. The next step is to formally verify in a theorem prover that the implementation implies all the properties extracted from the specifications.
Obviously, a mathematical
correlation must exist between the formal specification and the formal
model. The formalization of ray, wave, electromagnetic, and quantum optics play a vital role in both of these steps. Firstly, they provide the means to describe the specification and system model formally. Secondly, they also provide the formal reasoning support for verifying the system properties.

Finally, considering the mathematical complexity of optical system analysis, we may either encounter equations with no closed-form solution or problems in which the libraries developed in theorem prover are not rich enough to address them. CASs are the most efficient tools to provide solutions to such problems. Therefore, we propose to link our formal optical system analysis to a CAS, as shown in Figure \ref{fig:PF}. It is important to note here that this link would be used only for the cases where formal verification using a theorem prover is not an option. Obviously, other approaches, like numerical methods cannot compete with CASs in precision. Thus, as far as the whole analysis is concerned, the proposed method offers the most precise solution. We chose Mathematica to be the first CAS to be connected to our framework. This link will not only give us general access to Mathematica's symbolic algorithms but also to Optica \cite{OPTICA}, which is an optical design package for Mathematica.

Figure \ref{fig:PF} provides the formalization prerequisites (on the left side), along with the flow of theorem proving based analysis of optical systems (in the middle), and the connection between our approach and other tools (on the right side). The four libraries of ray, wave, electromagnetic, and quantum optics obviously share many concepts but are focused on different properties of optical systems. We have already formalized a significant portion of ray, electromagnetic, and quantum optics and we are currently working on the assessment of the preliminary steps to formalize wave optics.
Wave optics shares many
features of both ray and electromagnetic optics and hence can be formalized using either one of them. For
example, approximating electromagnetic fields under paraxial approximation or generalizing the
notion of a ray using wave functions essentially leads to the foundational concepts of wave optics \cite{Fund_of_Photonics}. In the next three sections of the paper, we present the existing HOL Light formalizations of ray, electromagnetic and quantum optics along with some insights on how to use them for analyzing optical systems.
Details of our formalizations and source codes can be find at \url{{http://hvg.ece.concordia.ca/projects/optics/}}.

\section{Ray Optics}\label{sec_ray}
\begin{figure}[h]
\centering
\scalebox{0.37}
{\includegraphics{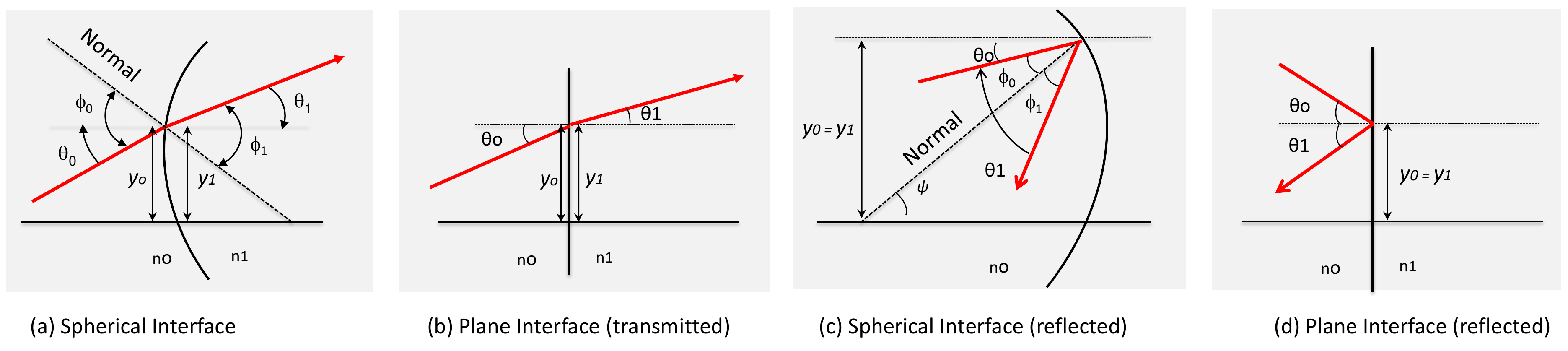}} \vspace{-.1cm}
\caption{Refraction and Reflection of a Ray}
\label{fig:interfaces} % labeling to refer it inside the text
\end{figure}

%\subsection{Overview}
\label{subsec:ray_overview}
Ray optics or geometrical optics characterizes light as rays and is  based on a set of postulates used to derive the rules for the propagation of light through an optical medium. These postulates are as follows\cite{Fund_of_Photonics}:
\begin{itemize}
\item Light travels in the form of rays emitted by a source,
\item An optical medium is characterized by its refractive index, and
\item Light rays follow Fermat's principle of least time.
\end{itemize}
Optical components, such as thin lenses, thick lenses, and prisms are usually centred about an optical axis, around which rays travel at small inclinations (angle with the optical axis). Such rays are called \textit{paraxial rays} and this assumption provides the
basis of \textit{paraxial optics}, which is the simplest framework of geometrical optics. The paraxial approximation explains how light propagates through a series of optical components and provides diffraction-free description of complex optical systems. When a ray passes through optical components, it undergoes  \textit{translation} or \textit{refraction}. In translation, the ray simply travels in a straight line from one component to the next and we only need to know the thickness of the translation. On the other hand, refraction takes place at the boundary of two regions with different refractive  indices  and the ray obeys the law of refraction, i.e., the angle of refraction relates to the angle of incidence  by the relation $n_{0}\phi_{0} = n_{1}\phi_{1}$, called  \textit{Paraxial Snell's law} \cite{Fund_of_Photonics}, where $n_{0}$, $n_{1}$ are the refractive indices of both regions and $\phi_{0}$, $\phi_{1}$ are the angles of the incident and refracted rays, respectively,  with  the normal to the surface. In order to model refraction, we thus need the normal to the refracting surface and the refractive indices of both regions. The refraction and reflection of a single ray from plane and spherical interfaces is shown in Figure \ref{fig:interfaces}.

The change in the position and inclination  of a paraxial ray as
it travels through an optical system can be efficiently described by the use of a
matrix algebra \cite{Lasers_66}. This matrix formalism (called \textit{ray-transfer matrices}) of geometrical optics provides accurate, scalable, and systematic analysis of real-world complex optical and laser systems. For example, we can relate the refracted and the incident ray for a spherical interface (Figure \ref{fig:interfaces} (a)) by a matrix relationship  as follows:

 \[
  \begin{bmatrix}
               y_{1}     \\[0.3em]
               \theta_{1}    \\[0.3em]

     \end{bmatrix} = \begin{bmatrix}
       1 &  0           \\[0.3em]
       \frac{n_{0}-n_{1}}{n_{1}R}  &  \frac{n_{0}}{n_{1}} \\[0.3em]
\end{bmatrix}
                    \begin{bmatrix}
              y_{0}     \\[0.3em]
              \theta_{0}    \\[0.3em]

     \end{bmatrix}
\]
 Finally, if we have an optical system consisting of $k$ optical components, then we can trace the input ray $R_{i}$ through all  optical components using composition of matrices of each optical component as follows:
\begin{equation}\label{EQ:op_system_eq}
 R_{o} = (M_{k}. M_{k-1}....M_{1}).R_{i}
\end{equation}
\noindent Simply, we can write $ R_{o} = M_{s}R_{i}$ where $ M_{s}=\prod_{i=k}^1 M_{i}$. Here, $ R_{o}$ is the output ray and $ R_{i}$ is the input ray. Note that the elements of ray-transfer matrices can be either real in case of spatial domain analysis or complex in case of time-domain analysis \cite{TimeDomain_ABCS_1998}.

\begin{figure}[h]
\centering
\scalebox{0.47}
{\includegraphics{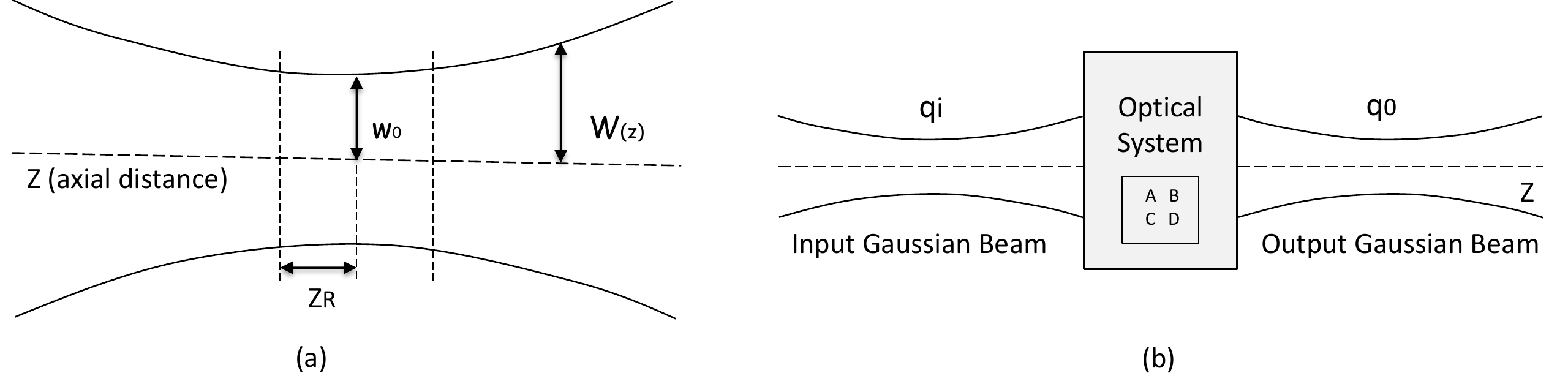}}\vspace{-.1cm}
\caption{(a) Gaussian Beam (b) Beam Transformation thorough an Optical System}
\label{fig:gaussian} % labeling to refer it inside the text
\end{figure}

Typical applications of ray-transfer matrices  are the stability analysis of  optical resonators \cite{Resonator_Stability_2011}, mode-locking, optical pulse transmission \cite{TimeDomain_ABCS_1998}, and  analysis of micro opto-electro-mechanical systems \cite{MOEMS_NASA}. Although ray tracing is a powerful tool for the early analysis of many optical systems, it cannot handle many situations due to the abstract nature of rays. For example, in laser applications, it is important to consider light as a \textit{beam} that provides more information than a simple ray. In most of the applications, such a beam of light is characterized by a \textit{Gaussian beam} \cite{Fund_of_Photonics}. In optics literature, there are different ways to model a Gaussian beam but one of the most common and effective way is the use of q-parameter, which is given as follows:
 \begin{equation}\label{}
   \frac{1}{q(z)} = \frac{1}{R(z)} - j \frac{\lambda}{\pi \omega^{2}(z)}
 \end{equation}
\noindent where $R(z)$  = $z[1 + \frac{z_{R}}{z}]$ is  the radius of curvature of the beam's wavefronts, $\omega(z)$ = $\omega_{0}[1 + (\frac{z_{R}}{z})^{2}]^{\frac{1}{2}}$  is the radius at which the field amplitude and intensity drop to $\frac{1}{e}$ and $\frac{1}{e^{2}}$ of their axial values, respectively. Note that $e$ represents the base of natural logarithm, $\omega_{0} = \omega(0)$ and $z$ represents the axial distance (see Figure \ref{fig:gaussian} (a)).
Similar to the ray-transfer matrix approach where each component is characterized by its matrix, another important aspect is to determine the output beam parameters corresponding to the input beam. This can be described by a well-known ABCD law of beam transformation for each optical component  and hence for the whole optical system using the elements of ray-transfer matrix of corresponding optical component as shown in Figure \ref{fig:gaussian} (b). Mathematically, the ABCD law is given as follows:
  \begin{equation}\label{}
   q_{o} =\frac{A. q_{i} + B} {C. q_{i} + D}
 \end{equation}
 \noindent where $q_{i}$ and $q_{o}$ represent the input and output beam q-parameters, respectively. The elements A, B, C, and D correspond to the final ray transfer matrix of an optical system.

The main applications of beam transformation are in the analysis of laser cavities, quasi optical systems, telescopes and  the prediction  of design parameters for physical experiments, e.g., recent dispersion-managed soliton transmission experiment \cite{Physica_review_ABCD}. In the next section, we present the complete formalization flow to encode the above mentioned fundamentals of ray optics to be able to formally reason about different aspects of optical and laser systems.

 \subsection{Formal Analysis Methodology}
\begin{figure}[h]
\centering
\includegraphics[width=4.9in]{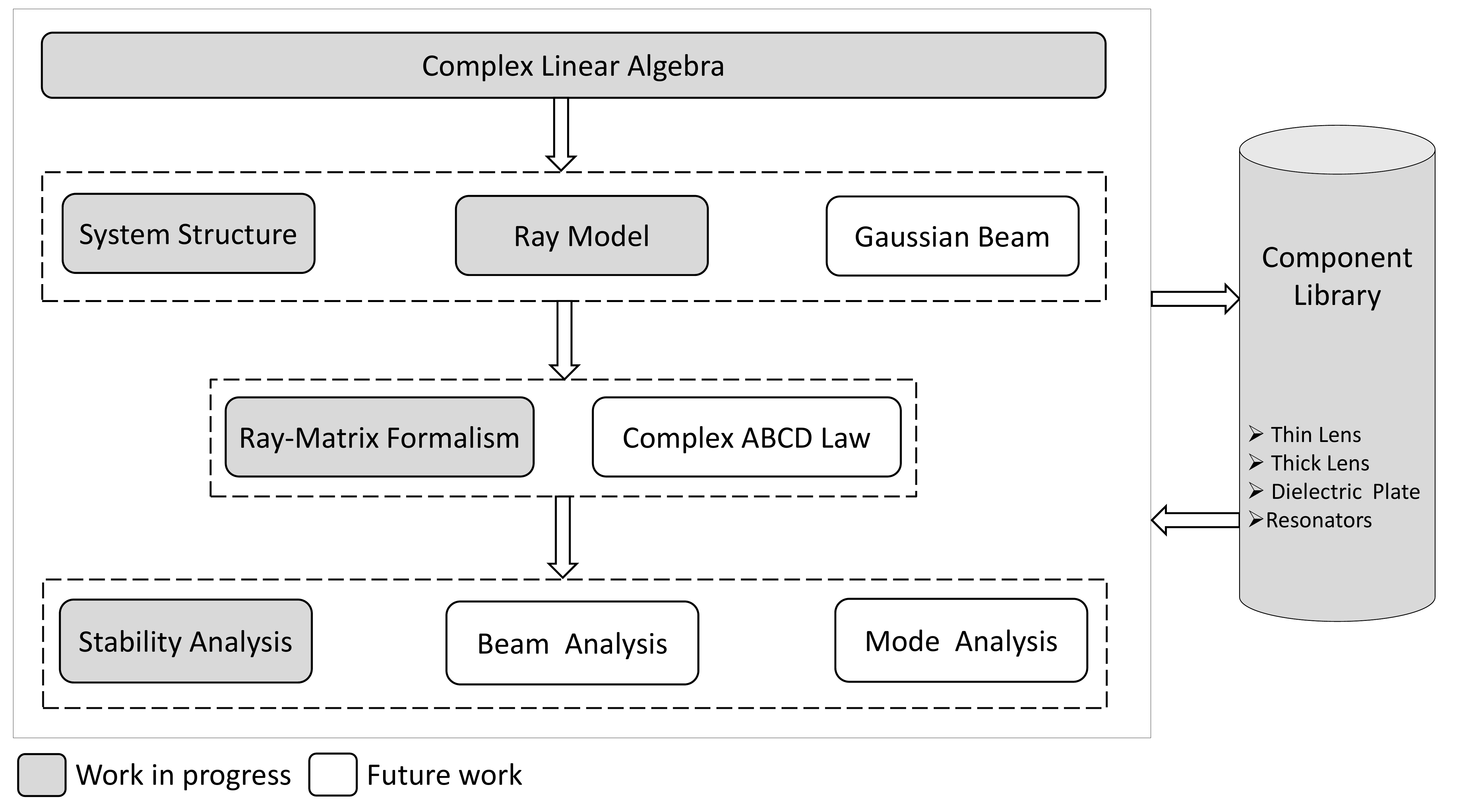}
\vspace{-.1cm}
\caption{Ray Optics Formalization Methodology}
\label{fig:flow} % labeling to refer it inside the text
\end{figure}

The proposed framework for the ray optics formalization, given in Figure \ref{fig:flow}, outlines the
necessary steps to encode theoretical  fundamentals of ray optics into a theorem prover.
The whole framework can be decomposed into four layers: first, the formalization of some complex linear algebra concepts, such as complex matrices and eigen-values, second, formalization related to the modelling of optical systems structure, modelling of rays and Gaussian beams, third, formalization related to system modelling, which are ray-transfer matrices and complex ABCD law, and finally, the properties of optical systems, such as stability, mode and output beam analysis.

 The first step in formal analysis is to construct a formal model of
the given system  in higher-order-logic based
on the description of the optical system and specification,
i.e., the spatial organization of various components and their parameters (e.g., radius of curvature of mirrors and distance between the components, etc.).
 In order to facilitate this step, we require the formalization of optical system structures, which consists of definitions of  optical interfaces (e.g., plane or spherical) and optical components (e.g., lenses and mirrors). The second step in the proposed framework  is  the formalization of the physical concepts of ray and Gaussian beams. Building on these fundamentals, the next step is to derive the matrix model of the optical system, which is basically a multiplication of the matrix models of individual optical components as described in Section \ref{subsec:ray_overview}.  This step also includes the formalization of the complex ABCD law of geometrical optics, which describes the relation between the input and output Gaussian beam parameters. At this point, the proposed framework provides all the fundamentals to model an optical system in a theorem prover.

 In order to facilitate the formal modelling of the system properties and reasoning about their satisfaction in the given system model, the next step is to provide the ability  to express system properties, i.e., their formal definitions and most frequently used theorems. These system properties are
  \emph{stability}, which ensures the confinement of rays within the system, \emph{beam analysis}, which provides the basis to derive the suitable parameters of Gaussian beams for a given system structure and  \emph{mode analysis}, which is necessary to evaluate the field distributions inside  the optical system when light traverses through that system. Finally, we apply the above mentioned steps to develop a library of frequently used optical components, such as lenses and mirrors. Since such components are the basic blocks of optical systems, this library helps to formalize new optical systems.

Next, we provide the highlights of the current status of our formalization related to some blocks (i.e., System Structure, Ray, Matrix Model and Stability) of Figure \ref{fig:flow}.

\subsection{HOL Light Implementation}
The formalization consists of three parts: 1) the formalization of  optical system structure; 2) the modelling of ray behaviour; 3) the formal verification of ray-transfer matrix of optical systems.
 \subsubsection*{ Optical System Structure}  Ray optics explains the behaviour of light when it passes through the free space and interacts with different interfaces, like spherical and plane, as described in Section \ref{subsec:ray_overview}.
   We can model free space by a pair of real numbers ($n,d$), which are essentially the refractive index and the total width in which ray can travel in free space. We consider only two fundamental interfaces, i.e., plane and spherical, which are further categorized as either transmitted or reflected.
Furthermore, a spherical interface can be described by its radius of curvature ($R$). We translate the above description in the HOL Light by defining some new types as follows\footnote{ From now on, all HOL Light statements will be written by mixing HOL Light script notations and pure mathematical notations in order to improve readability. Also, $\real$ and $\complex$ indicate the types real and complex, respectively.}:

\vspace{3pt}
 \begin{flushleft}
\begin{mdframed}
  \begin{definition}[Optical Interface and Free Space]\vspace{1pt} \ \\
\shol{\vdashdef  (free\_space \ =\real \times \real)}\vspace{1pt} \\
\shol{\vdashdef   \bitt{optical\_interface}\ =  plane\ |\ spherical\ \real} \vspace{1pt}\\
 \shol{\vdashdef  \bitt{interface\_kind\ }= transmitted\ |\ reflected}
  \end{definition}
\end{mdframed}
 \end{flushleft}

   An optical component is made of a free space (\hol{free\_space}) and an optical interface \\
   (\hol{optical\_interface}) as defined above.
   Finally, an optical system is a list of optical components followed by a free space. When passing through an interface, the ray is either transmitted or reflected (as shown in Figure \ref{fig:interfaces} (a-d)). In our formalization, this information is also provided in the type of optical components, as shown by the use of the type \shol{interface\_kind} as follows:

\vspace{3pt}
 \begin{flushleft}
\begin{mdframed}
 \begin{definition}[Optical Component and System] \vspace{1pt} $\mathtt{}$\\
\shol{\vdashdef (optical\_component:free\_space \times optical\_interface \times interface\_kind)}\\
\shol{\vdashdef (optical\_system:optical\_component\ list \times free\_space)}
  \end{definition}
\end{mdframed}
 \end{flushleft}

 \noindent   Note that this datatype can easily be extended to many other optical components if needed.

  A value of  type \shol{free\_space} does represent a real space only if  the refractive index is greater than zero. % (??Answered.?REVIEW 1-7).
  In addition, in order to have a fixed order in the representation of an optical system, we  impose that the distance of an optical interface
  relative to the previous interface is  greater or equal to zero. Next we assert  the validity of a value of type \shol{optical\_interface} by ensuring that the radius of curvature of spherical interfaces is never equal to zero.
  This yields the following predicates:

\vspace{3pt}
\begin{flushleft} \begin{mdframed}
  \begin{definition} [Valid Free Space and Valid Optical Interface]
    \label{def:valid_free_space} \vspace{1pt} $\mathtt{}$\\
	\shol{\vdashdef is\_valid\_free\_space ((n,d):free\_space)\ \Leftrightarrow\ 0 < n\ \wedge\ 0 \leq\ d}\\
	\shol{\vdashdef (is\_valid\_interface\ plane\ \Leftrightarrow\ T)\ \wedge\ }\vspace{1pt} $\mathtt{}$ \\
	\shol{\hspace{.8cm} (is\_valid\_interface\ (spherical\ R)\ \Leftrightarrow\  0 <> R)}
  \end{definition}
\end{mdframed} \end{flushleft}

  Then, by ensuring that this predicate holds for every component of an optical system, we can characterize valid optical systems (more details can be found in \cite{ADG_LNAI_UMAIR}).

\subsubsection*{Ray Model}
  We can now formalize the physical behaviour of a ray when it passes through an optical system.
  We only model the points where it hits an optical interface (instead of modelling all the points constituting the ray).
  So it is sufficient to just provide the distance of each one of these hitting points to the axis and the angle taken by the ray at these points.
  Consequently, we should have a list of such pairs $(distance,angle)$ for every component of a system.
  In addition, the same information should be provided for the source of the ray. For the sake of simplicity, we define a type for a pair $(distance,angle)$ as \shol{ray\_at\_point}. This yields the following definition:
  \vspace{3pt}
\begin{flushleft} \begin{mdframed}
 \begin{definition}[Ray]
    \vspace{1pt} $\mathtt{}$\\
	\shol{\vdashdef (ray\_at\_point :\real \times \real)}  \vspace{1pt}  \\
	\shol{\vdashdef (ray :ray\_at\_point \times ray\_at\_point \times (ray\_at\_point \times ray\_at\_point)\ list)}
  \end{definition}
\end{mdframed} \end{flushleft}

  The first \shol{ray\_at\_point} is the pair $(distance,angle)$ for the source of the ray, the second one is the one after the first free space,
  and the list of \shol{ray\_at\_point} pairs represents the same information for the interfaces and free spaces at every  hitting point of an optical system. Once again, we can specify what is a valid ray by defining some predicates, when it travels in free space and interacts with optical interfaces (again, for the sake of simplicity details are omitted and can be found in \cite{ADG_LNAI_UMAIR}).
 \begin{flushleft}
  \textbf{Verification of Ray-Transfer-Matrix of Optical Systems}
 \end{flushleft}

 We prove the generalized ray-transfer-matrix relation (\ref{EQ:op_system_eq}), which is valid for any ray and optical systems as follows:
\begin{flushleft} \begin{mdframed}
\begin{theorem}[Ray-Transfer-Matrix for Optical System]
\label{TH:ray-matrix-thm}  \vspace{1pt} $\mathtt{}$\\
	\shol{\hspace{.2cm}\vdash \Forall{sys, genray}}\\
	\shol{\hspace{.5cm} is\_valid\_optical\_system\ sys\ \wedge\ is\_valid\_ray\_in\_system\ genray\ sys\  \Rightarrow}  \vspace{1pt}  \\
	\shol{\hspace{.5cm}let\ (y_0,\theta_0),(y_1,\theta_1),rs = genray\ in } \vspace{1pt}  \\
	\shol{\hspace{.5cm}let\ y_n,\theta_n = last\_ray\_at\_point\ genray\ in  } \vspace{1pt}  \\
	\shol{\hspace{.5cm}\left[ \begin{array}{c} \mathtt{y_n} \\ \mathtt{\theta_n} \end{array}\right] =  system\_composition\ sys\ **\ \left[ \begin{array}{c} \mathtt{y_0} \\ \mathtt{\theta_0} \end{array} \right]}
	
  \end{theorem}
\end{mdframed} \end{flushleft}
\noindent Here, the parameters \shol{sys} and \shol{genray} represent the optical system and the ray, respectively. The function \shol{system\_composition} takes an optical system and returns the composition of matrices of optical components.  \shol{last\_ray\_at\_point} returns the last \shol{ray\_at\_point} of the ray in the system. Both assumptions in the above theorem  ensure the validity of the optical system and the good behaviour of the ray in  the system. The theorem is easily proved by  induction on the length of the system and by using previous results and definitions.

\section{Electromagnetic Optics}\label{sec_emf}

%\subsection{Overview}
\label{o_emf}

In the electromagnetic theory, light is described by the same principles that govern all forms of electromagnetic radiations.
An electromagnetic radiation is composed of an electric and a magnetic field.
The general definition of a field is ``a physical quantity associated with each point of space-time''.
Considering electromagnetic fields (``EMF''), the ``physical quantity'' consists of a 3-dimensional vector for the electric and the magnetic field.
Consequently, both those fields are defined as vector functions $\vec{E}(\vec{r},t)$ and $\vec{H}(\vec{r},t)$, respectively,
\noindent where $\vec r$ is the position and $t$ is the time.
These functions are related by the well-known Maxwell equations \cite{Born_99}:

\begin{equation}  \label{Maxwell equations}
\begin{array}{cp{.5cm}cp{.5cm}c}
\begin{aligned}
 \nabla\times\vec{E} &=-\Q{\vec{B}}{t} && \textnormal{,} &&
 \nabla\times\vec{H} =\vec{J}+\Q{\vec{D}}{t}\\
 \nabla\cdot\vec{D} &=\rho && \textnormal{,} &&
 \nabla\cdot\vec{B} =0 \\
\end{aligned}
\end{array}
\end{equation}

\noindent with their associated constitutive equations
%\vspace{-3mm}

\begin{equation}  \label{constitutive equations}
\begin{array}{cp{.5cm}cp{.5cm}c} %\vspace{-3mm}
\vec{D}=\varepsilon_0\vec{E}+\vec{P}=\varepsilon \vec{E} && \textnormal{and} && \vec{B}=\mu_0(\vec{H}+\vec{M})=\mu \vec{H} \vspace{-.1mm} \end{array}
\end{equation}

\noindent where $\vec{D}$ and $\vec{B}$ are the electric and magnetic flux density, respectively,  $\vec{J}$  the electric current density,
$\rho$ the electric charge density, and $\nabla\times$ and $\nabla\cdot$ denote the curl operation and divergence, respectively. The parameters $\varepsilon$ and $\mu$ represent the permittivity and permeability in the medium, and $\varepsilon_0$ and $\mu_0$ are permittivity and permeability in free space, respectively.
The vector fields $\vec{P}$ and  $\vec{M}$ represent the polarization and the magnetization density,
which are measures of the response of the medium to the electric and magnetic fields, respectively \cite{Pollock_95}.
Once the medium is known, an
equation relating $\vec{P}$ and $\vec{E}$, and another relating
$\vec{M}$ and $\vec{H}$ is established. When substituted in Maxwell
equations, the set of partial differential Equations (\ref{Maxwell
equations}) will be simplified governing only the two vector fields
$\vec{E}$ and $\vec{H}$. Therefore, to describe electromagnetic
waves in a medium, it would be enough to describe the \emph{medium},
and the \emph{electromagnetic fields} $\vec{E}$ and $\vec{H}$.

\subsubsection*{Medium Equations}

\begin{figure}
[!htb]
\begin {center}
 \includegraphics[width=2.6in]{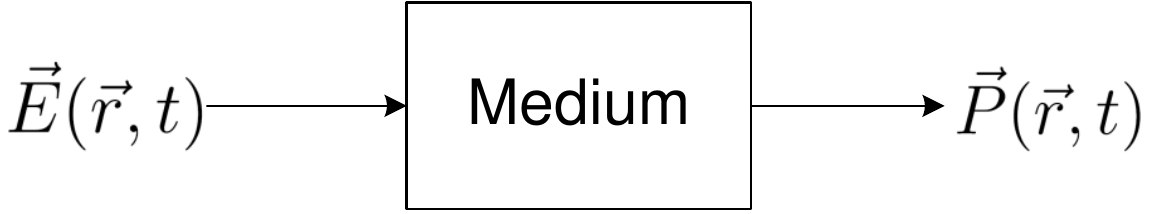}\vspace{-.2cm}
 \caption{Dielectric Medium Acting as a Filter}
 \label{medium_filter}
 \end {center}
\end{figure}

In most cases, mediums are considered to be non-magnetic, which
results in $\vec{M} = \vec{0}$ in Equation
(\ref{constitutive equations}). Consequently, the nature of the
dielectric medium is exhibited by the relationship between $\vec{P}$ and
$\vec{E}$, called medium equation. A very nice interpretation of
medium equation \cite{Saleh_91} is to consider medium as a filter
with electric field $\vec{E}$ as its input and polarization
density $\vec{P}$ as its output, shown in Figure
\ref{medium_filter}. $\vec{P}$ and $\vec{E}$ are both functions of
position, $\vec{r}$, and time, $t$.
A very famous and widely used model of medium is when it is considered to be linear, nondispersive, spatially
nondispersive, homogenous, and isotropic. In this case, the vectors
$\vec{P}(\vec{r},t)$ and $\vec{E}(\vec{r},t)$ are parallel and
proportional at any time and any position, and the medium equation can be described as follows:

\begin{equation*} \label{linear_ medium}
\forall \vec{r}, t \ \  \  \Rightarrow \ \ \ \  \vec{P} (\vec{r},t)= \varepsilon_0 \chi \vec{E}(\vec{r},t)
\end{equation*}

\noindent where $\chi$ is the electric susceptibility, which in this
case, is scalar, and is directly proportional to permittivity
$\varepsilon$.
In the absence of electric and magnetic sources, the Maxwell equations
describing such medium are simplified as follows:

\begin{equation}  \label{LME}
\begin{array}{cp{.5cm}cp{.5cm}c}
\begin{aligned}
 \nabla\times\vec{E} &=-\mu_0\Q{\vec{H}}{t} && \textnormal{,} &&
 \nabla\times\vec{H} =\varepsilon \Q{\vec{E}}{t}\\
 \nabla\cdot\vec{D} &=0 && \textnormal{,} &&
 \nabla\cdot\vec{B} =0 \\
\end{aligned}
\end{array}
\end{equation}

The set of Equations (\ref{LME}) will result in the famous Wave Equation
(\ref{wave_equation}), where $\vec{U}$ represents
any of the two fields $\vec{E}$ and $\vec{H}$
and $c$ is the speed of light in the medium.

\begin{equation} \label{wave_equation}
\nabla^2 \vec{U} - \frac{1}{c^2} \QQ{\vec{U}}{t} = 0
\end{equation}

\begin{table}
[!h]
\renewcommand{\arraystretch}{1.3}
\caption{Partial Differential Equations of Nonlinear, Spatially
Dispersive Media} \label{T_nonlinear_me}
\begin{center}
{\footnotesize
\begin{tabular}{|c|c|} \hline
{\bfseries Properties of the Medium} & {\bfseries Partial Differential Equation}  \\
\hline \hline
            Dispersive, Inhomogenous,  and Anisotropic
            & $ \nabla(\nabla\cdot\vec{E})-\nabla^2\vec{E} = - \varepsilon_0 \mu_0 \QQ{\vec{E}}{t} - \mu_0\QQ{\vec{P}}{t}$  \\ \hline
            Dispersive, Homogenous,  and Isotropic
            & $\nabla^2 \vec{E} - \frac{1}{c^2} \QQ{\vec{E}}{t} = \mu_0\QQ{\vec{P}}{t} $  \\ \hline
            Nondispersive, Homogenous,  and Isotropic
            & $\nabla^2 \vec{E} - \frac{1}{c^2} \QQ{\vec{E}}{t} = \mu_0\QQ{f(\vec{E})}{t} $  \\ \hline
  \end{tabular}}
\end{center}
\end{table}

Table \ref{T_nonlinear_me} refers to the three
cases, which are extensively used in optical device analysis. As it can be observed, the system equations are expressed by
partial differentiations, and depending on the application, the system model describing the behaviour of medium can become very  complex, with no closed-form solution.
\subsubsection*{Electromagnetic Fields} \label{fields}

An electromagnetic wave can be considered as monochromatic or
polychromatic.
Any polychromatic electromagnetic wave can be
considered to be composed of monochromatic components. When the wave
light is monochromatic, all the components of the electric and
magnetic fields are harmonic functions of time of the same
frequency. In this case, electromagnetic fields are expressed in
terms of their complex amplitudes, $\vec{U}(\vec{r})$, where
$\vec{U}$ can be either electric field $\vec{E}$ or magnetic field
$\vec{H}$.

\begin{equation} \label{monochromatic waves}
\begin{aligned}
\vec{U}(\vec{r},t) &= \vec{U} (\vec{r}) e^{j\omega t} \\
\vec{U}(\vec{r})  &= a(\vec{r}) e^{j\phi(\vec{r})}
\end{aligned}
\end{equation}

At a given position $\vec{r}$, $\Vec{U}(\Vec{r})$ is called complex
amplitude, which is defined by a complex variable with magnitude,
$a(\vec{r})$, which is the amplitude of the field and with argument,
$\phi (\vec{r})$, which is the phase of the field.

Depending on the waveform, different solutions can be considered for
the monochromatic waves. The simplest and most important solution is the plane wave,
which is defined based on its wavefronts. Wavefronts are defined as
surfaces of equal phases, which means $\phi(\vec{r})$ is constant
for all $\vec{r}$.
A plane wave is a constant-frequency wave for which the wavefronts are
infinite parallel planes of constant amplitude. The complex
amplitude of plane wave is defined as:

\begin{equation} \label{plane waves}
\begin{aligned}
\vec{U}(\vec{r})  &= A e^{-j\vec{k}\cdot \vec{r}}
\end{aligned}
\end{equation}

\noindent where $A$ is a complex constant called complex envelope
and $\vec{k} = (k_x, k_y, k_z)$ is called wavevector, the magnitude
of the wavevector, $\vec{k}$, is called wavenumber $k$, and is
correlated to the wavelength $\lambda$ and consequently to the frequency
$\nu$.

\begin{equation} \label{lambda}
\begin{aligned}
\lambda  &= \frac{2 \pi}{k} = \frac{c}{\nu} \\
\end{aligned}
\end{equation}

It can be shown that the intensity of plane wave, which is defined
as the optical power per unit area, is constant everywhere in space,
which means plane waves are idealized models. However, in practice,
light waves are frequently approximated as plane waves in a
localized region of space.

The next waveform, which is used as an approximation to real
waveforms, is the paraxial wave. Paraxial waves can be considered as an
extension to plane waves. One way to describe paraxial wave is to
have a plane wave $A e^{-j\vec{k} \cdot \vec{r}}$, where its complex
envelope $A$, is slowly varying in space. Hence, a paraxial wave can be
described as:

\begin{equation} \label{paraxial waves}
\begin{aligned}
\vec{U}(\vec{r})  &= A(\vec{r}) e^{-j\vec{k}\cdot\vec{r}} \\
\end{aligned}
\end{equation}

The variation of $A(\vec{r})$ should be slow within the distance of
a wavelength $A = \frac{2\pi}{k}$, so that the wave approximately
maintains its plane wave nature. Paraxial waves are also not exact
solutions of monochromatic waves, but an approximation, that is
widely used in Optics.

As it can be observed from Equations (\ref{monochromatic waves}),
(\ref{plane waves}), (\ref{lambda}), and (\ref{paraxial waves}), to
formally define and reason over electromagnetic waves, complex
vectors and vector operations, like dot product, are needed.
In the next section, we present the complete formalization flow to encode the above mentioned fundamentals of electromagnetic optics to be able to formally reason about different aspects of optical systems based on this very rich theory.

\subsection{Formal Analysis Methodology}\label{sec_methodology}
\begin{figure}[!hbt]
\centering
\includegraphics[width=4.9in]{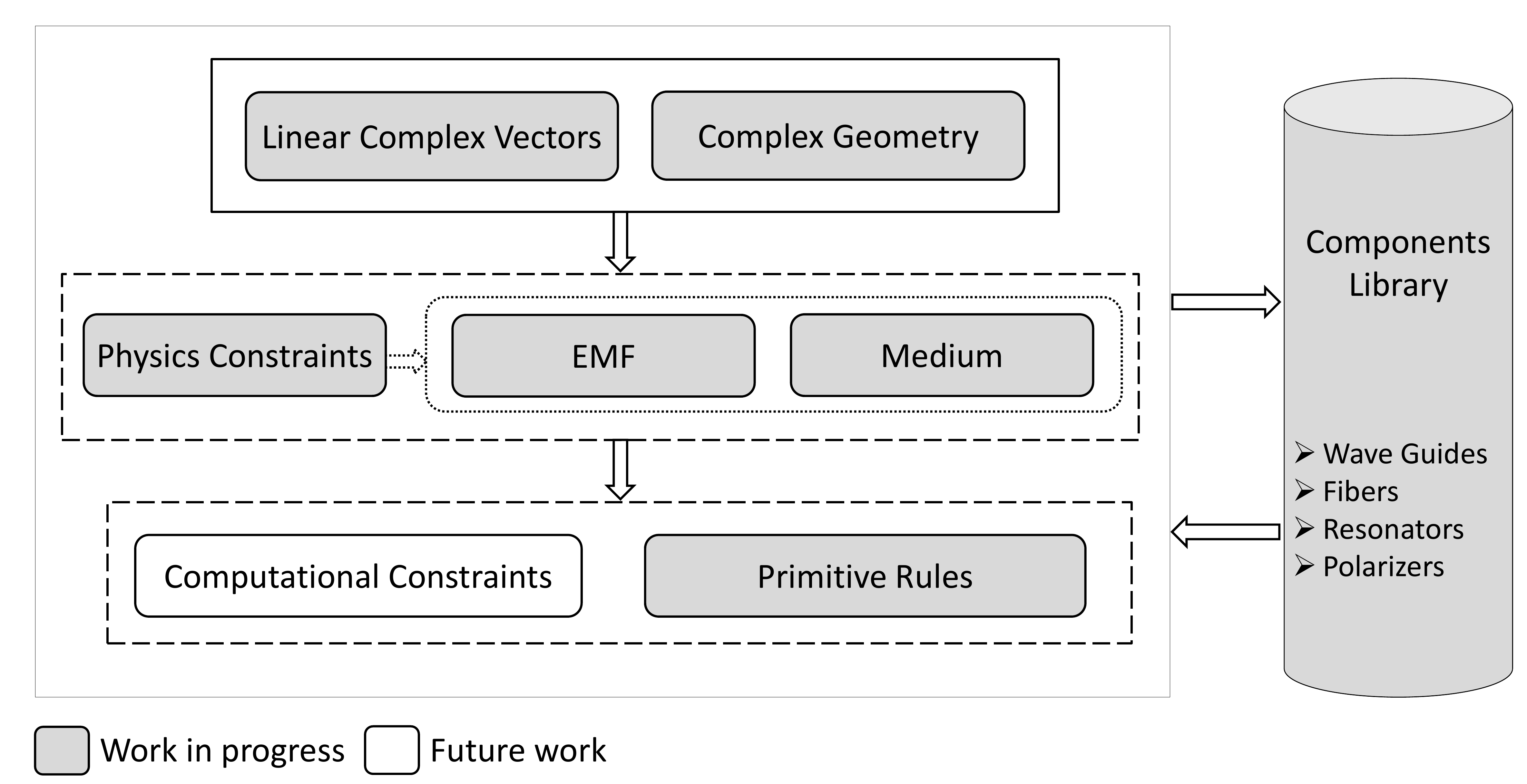} \vspace{-.1cm}
 \caption{Electromagnetic Optics Formalization Methodology}
 \label{pf_emf}
 \end{figure}

Figure \ref{pf_emf} shows a framework of formal verification of optical systems based on the electromagnetic theory.
The EMF aspects of an optical system are usually described mathematically using complex vectors. Whereas the medium aspects are mathematically expressed using Euclidean and non-Euclidean geometry and complex calculus.
Just like ray optics, the first essential block that needs to be developed is the libraries on complex calculus, geometry and other mathematical concepts, which are necessary to describe EMFs and mediums. Obviously not all the mathematical concepts are formalized and there is a dire need to improve existing theories on multivariate calculus to build a strong infrastructure to reason over optics. Our major contribution in this part is the development of a rich library on complex vector analysis \cite{sanaz} and complex geometry.

 The third block in the second layer of Figure \ref{pf_emf} is dedicated to the library of constraints. To model a system based on electromagnetic optics, in practice, different sets of assumptions are required, in order to simplify the Maxwell equations. For example, EMFs are considered as plane waves, or mediums are considered to be linear and homogeneous. These assumptions are enforced by modelling the system by physicists and optical engineers.

All aforementioned blocks, so far, are necessary to formally describe the system model and its specifications, which are indicated as formal model and formal specification in Figure \ref{fig:PF}. The next step in our formal analysis is to formally express that the system satisfies, or implies, its required specification. This implication has to be verified within the sound core of a theorem prover. This can of course be done from scratch by using only the inference rules of higher-order logic. But some fundamental results are always used, irrespective of the optical component that we want to verify, e.g., the law of Reflection, Snell's law, or Fresnel equations.
So we propose to prove these foundations once and for all in order to make the verification of new components easier.
This yields a ``library of primitive rules of optics''.

Optical systems are usually composed of some commonly used sub-systems, like resonators or waveguides. Therefore, we also propose to formalize such commonly-used structures so that complex optical systems can be modelled and analyzed easily in a hierarchical manner. The fact that our formalization starts from the low-level roots of optics not only allows us to formalize these commonly-used structures, but also provides the ability to define new structures when needed.
Note that new formalized structures can be added to the library of components and sub-systems in order to be used without enduring the pain of formalizing them again.

Finally, considering the mathematical complexity of optical system analysis, we may encounter equations with no symbolic (or ``closed-form'') solution.
We will explain in Section \ref{sec_CAS}, in more details how we are connecting HOL Light with Mathematica to fulfil our requirements.
Obviously, this connection has the risk of error due to the complex
algorithms used in the core of CAS \cite{harrison_thesis}.
Thus, it is recommended to verify the answers derived by CAS within HOL Light. Obviously, this approach is not always possible, specially when the simplifications involve numerical approximations.
We can trust the CAS and tag those theorems proved in a hybrid fashion, by the name of CAS as proposed in
\cite{Harrisson}. However, we would like to include as much details as possible within these tags. These tags are producing the last block of our framework called Computational Constraint.

In the next section, we  provide the highlights of the current status of our formalization related to those blocks in Figure \ref{pf_emf}, which are directly related to the concept of electromagnetic optics (i.e., EMF, Medium, Physics Constraints, and Primitive Rules).

 \subsection{HOL Light Implementation}
\label{sec_formalization}
In this section, to show the flow of our proposed framework, we provide the higher-order-logic formalization of the electromagnetic model of light wave and the formal verification of some primitive laws of optics.
As explained above, the electromagnetic theory considers light as an electromagnetic field (``EMF'').
Thus, we first need to define a field. The general definition of a field is ``a physical quantity associated with each point of space-time''.
Points of space are represented by 3-dimensional real vectors, so we define the type \hol{point} as an abbreviation for the type \hol{real^3}
(\hol{A^N} is the HOL Light library built-in type for vectors of size \hol N\ whose components are of type \hol A \cite{harrison_05}).
Also, time is represented by a real number. Again, we define the type \hol{time} as an abbreviation for the type \hol{real}.
Finally, the ``physical quantity'' is formally defined as a 3-dimensional complex vector.
Consequently, the type \hol{field} (either magnetic or electric) is defined as \hol{point \to time \to complex^3}.
Then, since an EMF is composed of an electric and a magnetic field, we define the type \hol \emf\ to represent \hol{point \to time \to complex^3 \times complex^3}.

A very general expression of an EMF is
$\vec{U}(\vec{r},t) = \vec a(\vec{r}) e^{j\phi(\vec{r})} e^{j\omega t}$,
\noindent where $\vec{U}$ can be either the electric or magnetic field at point $\vec{r}$ and time $t$.
We call $\vec a(\vec{r})$ the \emph{amplitude} of the field and $\phi (\vec{r})$ its \emph{phase}.
Note that we consider only monochromatic waves, with  \emph{frequency} $\omega$.

Here, we focus on monochromatic \emph{plane} waves, where the phase $\phi(\vec{r})$ has the form $-\vec{k} \vdot \vec{r}$, defined using the dot product between real vectors. We call $\vec k$ the \emph{wavevector} of the wave and it represents the propagation direction of the wave.
This yields the following definition:\\

 \vspace{5pt}
\begin{flushleft} \begin{mdframed}
 \begin{definition}[Plane Wave]  \vspace{1pt}
\label{def_planar}
\shol{\vdashdef\planewave (k:\real^3)\ (\omega:\real)\ (E:\complex^3)\ (H:\complex^3) :\ \emf}  \vspace{1pt}\\
 \shol{ \hspace{3cm}\eqdef \Lambda{(r:point)\ (t:time)} (e^{-j(k \vdot  r - \omega t)} E, e^{-j ( k\vdot  r - \omega t)} H)}
\end{definition}
%\vspace{-.5mm}
\end{mdframed} \end{flushleft}

\noindent where \hol{j} denotes $\sqrt{-1}$.
Note that, although complex numbers are already defined in HOL Light \cite{harrison_07}, we had to develop our own library of \emph{complex vectors} \cite{sanaz}
in order to define operations like addition, multiplication by a scalar or dot product for such vectors.
In addition to Definition \ref{def_planar}, we define the helper predicates and the functions
\hol\isplanewave\NS, \hol\kofwave\NS, \hol\wofwave\NS, \hol\eofwave\NS, and \hol\hofwave such that:

\begin{flushleft} \begin{mdframed}
\begin{constraint}[] \vspace{1pt}
\ \\
\shol{\Forall{emf}\isplanewave emf \Leftrightarrowdef } \vspace{1pt}\\
\shol{emf = \planewave(\kofwave emf) (\wofwave emf)(\eofwave emf)\ (\hofwave emf)}
\end{constraint}
%\vspace{-.5mm}
\end{mdframed} \end{flushleft}

\begin{figure}[!htb]
  \centering  % Requires \usepackage{graphicx}
  \includegraphics[width=2.8in]{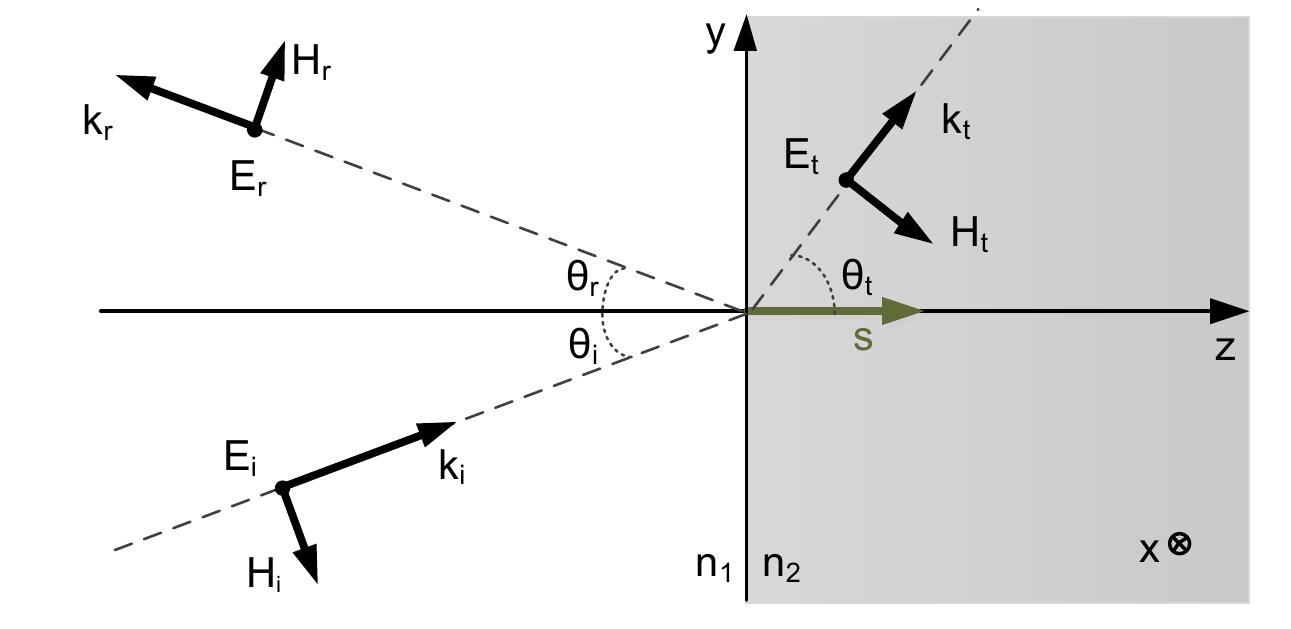}
  \caption{Plane Interface between Two Mediums}\label{fig_interface}
\end{figure}

When a light wave passes through a medium, its behaviour is governed by different characteristics of the medium.
The \emph{refractive index} is the most dominant among these characteristics and thus we have used the data type \hol{medium} to represent the medium with its refractive index, which is a real number.
Most of the study of an optical device deals with the passing of light from one medium to another.
So our basic system of study is the \emph{interface} between two mediums.
In general, such an interface can have any shape, but, most of the time, a plane interface is used, as demonstrated in Figure \ref{fig_interface}.

\vspace{7mm}
So we define the type \hol{interface} as \hol{medium \times medium \times plane \times real^3}, i.e., two mediums, a plane (defined as a set of points of space), and a orthonormal vector to the plane,
indicating which medium is on which side of the plane.

Another useful consequence of Maxwell equations is that the projection of the electric and magnetic fields shall be equal on both sides of the interface plane \cite{Pollock_95}.
This can be formally expressed by saying that the cross product between those fields and the normal to the surface shall be equal:

\vspace{5pt}
\begin{flushleft} \begin{mdframed}
 \begin{definition}[Boundary Conditions]
 \label{bound_cond}\ \\ \shol{\vdashdef \boundaryconds emf_1\ emf_2\ n\ p\ t \Leftrightarrowdef}  \vspace{1pt}\\
\shol{\hspace{.8cm} n\times\eofemf emf_1\ p\ t = n\times \eofemf emf_2\ p\ t\  \wedge\ }  \vspace{1pt} \\
\shol{\hspace{.8cm} n\times \hofemf emf_1\ p\ t = n\times \hofemf emf_2\ p\ t}
\end{definition}
\end{mdframed} \end{flushleft}
\vspace{5pt}

%
%\vspace{-.5mm}
\noindent where $\times$ denotes the \emph{complex} cross product,
and \hol\eofemf and \hol\hofemf are helper functions returning the electric and magnetic field components of an EMF, respectively.

Now that the notions of EMF and interface have been formalized, we
can prove some basic properties of optics that constitute the foundations to verify any optical system. In Figure \ref{pf_emf}, they are referred to as ``primitives''.
Most of these primitives impose some particular constraints on the waves, for instance, some parameters must be positive, or non-null.
One of the major advantages of theorem proving over other analytical methods is that these constraints are explicitly provided in the hypotheses of the corresponding theorems.
This way, these theorems can only be applied if the corresponding constraints are ensured.
As already explained, the study of an optical component mostly deals with the behaviour of light when it passes from one medium to another.
Thus, we first formalize the simple case of a plane interface between two mediums, in the presence of a plane wave, shown in Figure \ref{fig_interface}, with the following predicate:
\def\emfinmed{emf\_in\_med\ }

\vspace{3pt}
\begin{flushleft} \begin{mdframed}
 \begin{constraint}\label{is_plane_wave} \ \\
	\shol{\vdashdef \isplanewaveatitf i\ emf_i\ emf_r\ emf_t \Leftrightarrowdef} \vspace{1pt}\\
	\shol{\hspace{.8cm}\isvaliditf i\ \wedge\  \nonnull emf_i\  \wedge\ }  \vspace{1pt}\\
	\shol{ \hspace{.8cm}\isplanewave emf_i\ \wedge\ \isplanewave emf_r\ \wedge\  \isplanewave emf_t\ \wedge } \vspace{1pt}\\		
	\shol{\hspace{.8cm} (let\ (n_1,n_2,p, n) = i\ in\ }  \vspace{1pt}\\
	\shol{\hspace{1.1cm}\Forall {pt}\ \isinplan pt\ p \Rightarrow\ }  \vspace{1pt}\\
	\shol{\hspace{1.1cm} \Forall t \boundaryconditions (emf_i+emf_r)\ emf_t\ n\ pt\ t)\ \wedge} \vspace{1pt} \\
	\shol{\hspace{.8cm}( let\ (k_i,k_r,k_t) = \mapt \kofwave (emf_i,emf_r,emf_t)\ in}\\
	\shol{\hspace{1.1cm} 0\leq (k_i\vdot n)\ \wedge\ (k_r\vdot n)\leq0\ \wedge\ 0 \leq (k_t\vdot n)\ \wedge\ } \vspace{1pt} \\
	\shol{ \hspace{1.1cm}\Exists{k_0}\ norm\ k_i = k_0 n_1\ \wedge\
 norm\ k_r = k_0 n_1\ \wedge\ norm\ k_t = k_0 n_2)\ \wedge} \vspace{1pt}\\	
	\shol{\hspace{.8cm}let\ \emfinmed = \Lambda{emf\ n}\hofwave emf = \frac{1}{\eta_0  k_0}(\kofwave emf)\times(\eofwave emf)) in}\\
 	\shol{\hspace{1.1cm}\emfinmed emf_i\ n_1\ \wedge\  \emfinmed emf_r\ n_1\ \wedge\ \emfinmed emf_t\ n_2}
\end{constraint}%\vspace{-.5mm}
\end{mdframed} \end{flushleft}

\vspace{4pt}
\noindent where \hol{\mapt f\ (x,y,z)=(f\ x,f\ y,f\ z)} and \hol{\eta_0} is the impedance of vacuum, a physical constant relating magnitudes of electric and magnitude fields of electromagnetic radiation travelling through vacuum.
The predicate of Constraint \ref{is_plane_wave} takes an interface \hol i\ and three EMFs \hol{emf_i}, \hol{emf_r}, and \hol{emf_t},
intended to represent the incident wave, the reflected wave, and the transmitted wave, respectively.
When \hol\isplanewaveatitf holds, it first ensures that the arguments are wellformed,
i.e., \hol i\ is a valid interface and the three input fields are plane waves.
It also ensures that the reflected wave exists by asserting that its electric field is non-null (both electric and magnetic fields of an EMF are not null) and goes from medium 1 to medium 2, and that the reflected and transmitted waves go in the opposite and same direction, respectively.
These conditions are expressed by using the dot product of the wavevectors to the normal of the interface plane.
Moreover, Definition \ref{is_plane_wave} also ensures that the boundary conditions shall hold at every point of the interface plane and at all times.

From this predicate, which describes the interface in Figure \ref{fig_interface}, we can, immediately, reason over some geometrical properties of the wave; for instance, the law of plane of incidence which indicates the fact that the incident, reflected, and transmitted waves all lie in the same plane, called plane of incidence:

\begin{flushleft} \begin{mdframed}
%\vspace{-1mm}
 \begin{theorem}[Law of Plane of Incidence] \label{thm_incidence}
 \ \\
         \shol{\hspace{.2cm}\vdash \Forall{i\ emf_i\ emf_r\ emf_t}} \vspace{1pt}\\
         \shol{\hspace{.5cm}\isplanewaveatitf i\ emf_i\ emf_r\ emf_t\ \wedge\ }\vspace{1pt}\\
	\shol{\hspace{.5cm}\nonnull emf_r\ \wedge\  \nonnull emf_t\ \Rightarrow\ let \ n = \normalofint i\ in}\vspace{1pt}\\
	\shol{\hspace{.5cm}coplanar\ \{ vec\ 0,\kofwave emf_i,\kofwave emf_r,\kofwave emf_t, n\} }
\end{theorem}
\end{mdframed} \end{flushleft}

A second geometric consequence is the fact that the reflected wave is symmetric to the incident wave with respect to the normal to the surface:

\begin{flushleft} \begin{mdframed}
 \begin{theorem}[Law of Reflection] \label{thm_reflection}\ \\
         \shol{\hspace{.2cm}\vdash \Forall{i\ emf_i\ emf_r\ emf_t} }\vspace{1pt}\\
         \shol{\hspace{.5cm}\isplanewaveatitf i\ emf_i\ emf_r\ emf_t\ \wedge\ }\vspace{1pt}\\
	\shol{\hspace{.5cm}\nonnull emf_r \ \Rightarrow let\ n = \normalofint i\ in}\vspace{1pt}\\
	\shol{\hspace{.5cm}\aresymetric (-(\kofwave emf_i))\ (\kofwave emf_r)\ n}
\end{theorem}  %\vspace{-1mm}
 \end{mdframed} \end{flushleft}

\noindent where \hol{\aresymetric}\vectt u \vectt v \vectt w formalizes the fact that \vectt u and \vectt v are symmetric with respect to \vectt w
(this is easily expressed by saying that \texttt{\vectt v = 2$*$(\vectt u $\vdot$ \vectt w)\vectt w - \vectt u} ).
Referring to Figure \ref{fig_interface}, Theorem \ref{thm_reflection} just means that $\theta_i=\theta_r$, which is the expression usually found in optics literatures.

The formal proofs of the above theorems heavily rely upon complex vectors and multivariate transcendental functions properties. Obviously, their development is significantly harder than that of their informal counterparts, especially since
proofs in physics textbooks make many mathematical assumptions and simplifications that are not always justified, or are
justified only by \emph{physical} considerations without any \emph{mathematical} arguments.
The major advantage of these formalizations is the ability to utilize them to formally analyze optical systems.
We showed the effectiveness of our developed theories by the formal analysis of some optical components in which their properties are used in many practical applications. One example is formalization of resonant cavity, which is the building block of resonant cavity enhanced devices \cite{Unlu_95}. We formalized the quantum efficiency and optical power inside the resonant cavity. These two properties are of high interest in developing many applications, including photo-detectors \cite{karimi_13}, emitting lasers \cite{chen_12}, and fundamental structures like, vertical cavity surface emitting lasers (VCSEL) \cite{Unlu_95}, which are used in optical data communication, position sensing, biochemical sensing, and imaging applications \cite{meng_13}.

%%%%%
\section{Quantum Optics}\label{sec_quantum}

On the contrary to what we present in the previous two sections, quantum optics considers light as a  stream of particles called photons. This concept of photons reveals  new properties and phenomena about the light, especially at a low number of photons \cite{opticalcoherence}. Moreover, it allows a better use of existing optical devices, e.g., beam splitters \cite{simpledevices}, and the invention of totally new quantum devices, e.g., single photon devices \cite{Single_photon}. These devices help in various aspects such as  performance, e.g., detection of gravitational waves, and
sometimes provide novel solutions, e.g., quantum computation \cite{coherentcomputer}.

The verification of quantum systems maintains the same previously mentioned problems. Moreover, the computer simulation is not practical since Feynman proved that quantum systems
cannot be efficiently simulated on ordinary computers (it requires to solve an exponential number of differential equations) \cite{Simulatingphysics}. In such systems, \emph{physical lab simulation} is performed, which poses cost and safety problems \cite{opticlab}: every little optical element varies in cost from a few hundred to a few thousand dollars \cite{opticlab}. In addition, scientists and engineers who carry out the simulation process should be well protected against the beams due to their harmful nature \cite{opticsaft}. This clearly increases the importance of formal analysis technique in the area of quantum optics.

One of the essential applications of quantum optics is the provision of  some practical models  for implementing quantum computers. Such computers provide  promising solution in solving hard computational problems \cite{Jenn_11}. One of the vital goals of any of these optical models is to assure the satisfiability between the quantum computer mathematical specifications and the quantum optics model (or implementation). Tackling this task not only requires formalization of definitions and theorems, but also needs implementation of optical elements, such as a \emph{beam splitter}.

\bigskip
\noindent \textbf{Quantum State}. Any physical system has a state that describes the system dynamics at a particular time. Usually, it is formed by a set of system atomic information $\overrightarrow{x}$ (or coordinates), e.g., a position of a moving particle. Classically, we can deterministically define a system state at any time. On the other hand, in quantum theory, the evaluation of a system state possesses a probabilistic  notion, i.e., available information about the system are probabilities of being at specific states. Therefore, \emph{a quantum state} $\psi( \overrightarrow{x})$ is mathematically described as a probability density function (PDF) or more accurately, it is a complex-valued function and $\psi^{\ast}(\overrightarrow{x}) \psi(\overrightarrow{x})$ is a PDF Thus, a quantum state satisfies the PDF properties, in particular, square integrability:

\begin{equation}\label{eq:inegrable}
   \int_{-\infty}^{\infty} \psi^{\ast}(\overrightarrow{x}) \psi(\overrightarrow{x}) d \overrightarrow{x}=1
\end{equation}
\noindent If we collect all square integrable complex-value functions, we get the set of  quantum states (note that  this set changes according the type and number of the system coordinates). Actually, this set  forms an inner space with the integration as an inner product function \cite{Quantum_ref}, and this is the most important information in order to determine a  quantum state.

Now for a quantum state,  we are interested in the following properties from the formalization point of view:
\begin{itemize}
  \item Quantum state is a complex-valued function.
  \item Universal set of quantum states is an inner product  vector-space.
\end{itemize}

\bigskip
\noindent \textbf{Quantum Operator}. In general, physicists are interested in observable quantities besides the state of the system, such as the velocity of a moving particle.  Such kind of  information can be  derived from system atomic information, i.e., it is a function $\hat O(\overrightarrow{x})$. Due to such  nature of quantum systems, we are interested in the expectation of such observables:

 \begin{equation}\label{eq:cordexpec}
    E[\hat O] =  \int_{-\infty}^{\infty} \hat O\ \psi^{\ast}\ \psi d\overrightarrow{x}
 \end{equation}
 \noindent or equivalently:
   \begin{equation}\label{eq:cordexpec}
    E[\hat O] =  \int_{-\infty}^{\infty} \psi^{\ast}\ \hat O \ \psi d\overrightarrow{x}
 \end{equation}

 \noindent The above expression can be seen as an inner product between $\psi$ and $\hat O\ \psi$. Note that it is equivalent to the application of a  function $\hat O$ to the vector $\psi$, which results in a new vector. In addition to what is proved about quantum state, it also proved that an observable  $\hat O$ is a linear self-adjoint transformation over the quantum state space. From now on, we call such observables as  \emph{quantum operators}. In general, the Self-adjoint operator satisfies the following property:

 \begin{equation}\label{eq:selfadjoint}
     \int_{-\infty}^{\infty} \psi_1^{\ast}\ \hat O \ \psi_2 d\overrightarrow{x} = \int_{-\infty}^{\infty} \hat O\ \psi_1^{\ast}\  \psi_2
 \end{equation}

 Again, from the formalization point of view, we are interested in the following properties of quantum operators:

 \begin{itemize}
   \item Quantum operator is a function with a complex-valued function as its domain and range.
   \item Quantum operator is a linear self-adjoint transformation.
 \end{itemize}
\bigskip
\textbf{Quantum Optics}.
Many of physical systems studied in quantum mechanics were studied before in classical theory. This conversion (i.e., between classical and quantum) is commonly implemented using \emph{the canonical quantization} \cite{quantization}. Then, quantum optics comes as the result of the canonical quantization of light. In classical theory, light is an electromagnetic field (as presented before), which would be a single-mode field (i.e., single resonance frequency $\omega$) or multi-mode field. For simplicity, we will study the quantum model of a single-mode field since the obtained results  still apply to multi-mode with minor modifications \cite{opticalcoherence}.

In quantum theory, single-mode is described as follows:
\begin{enumerate}
  \item System coordinate atomic information: the charge density $\hat{q}$ and flux intensity  $\hat{p}$ inside the field. And  $[\hat{q},  \hat{p}] = \hat{q}\ \hat{p} - \hat{p}\ \hat{q}=i\hbar$ where $\hbar$ is the Planck's constant and $[\hat{q},  \hat{p}]$ is called a commutator. Note that commutator of coordinates is one of the canonical quantization postulates.
  \item The amount of energy inside the field:
  \begin{equation}\label{quhml}
    \hat{H} =  \frac{\omega^{2}}{2} \hat q^{2} + \frac{1}{2} \hat p^{2}
  \end{equation}
   Note that  $\hat{A}^{2}$ denotes $\hat{A}\circ\hat{A}$, i.e., for operators, the multiplication is actually the composition (which is not necessarily commutative).
 \end{enumerate}

  On the basis of above information, in addition to quantum postulates, many theories were developed that show advantages of quantum optics over classical models of light. The following are the common theories used in quantum system analysis, which are also in our interest for the formalization purposes:
 \begin{itemize}

   \item The \emph{multi-mode} field: this allows the study of multi-input multi-output devices (e.g., beam splitters).
  \item \emph{Light states}. As a quantum system, light has a quantum state, and we are interested in some special cases of it, in particular, \emph{coherent state} (e.g., laser sources) and  \emph{squeezed state}.

  \item \emph{Detection theory}. This theory is concerned with how we can detect each single photon of light and count the whole number of photons which is useful in quantum computation.
\end{itemize}

Next, we present our formal analysis methodology for quantum systems.

\subsection{Formal Analysis Methodology}\label{sec:form}

\begin{figure}[h]
 \centering
	\includegraphics[width=4.8in]{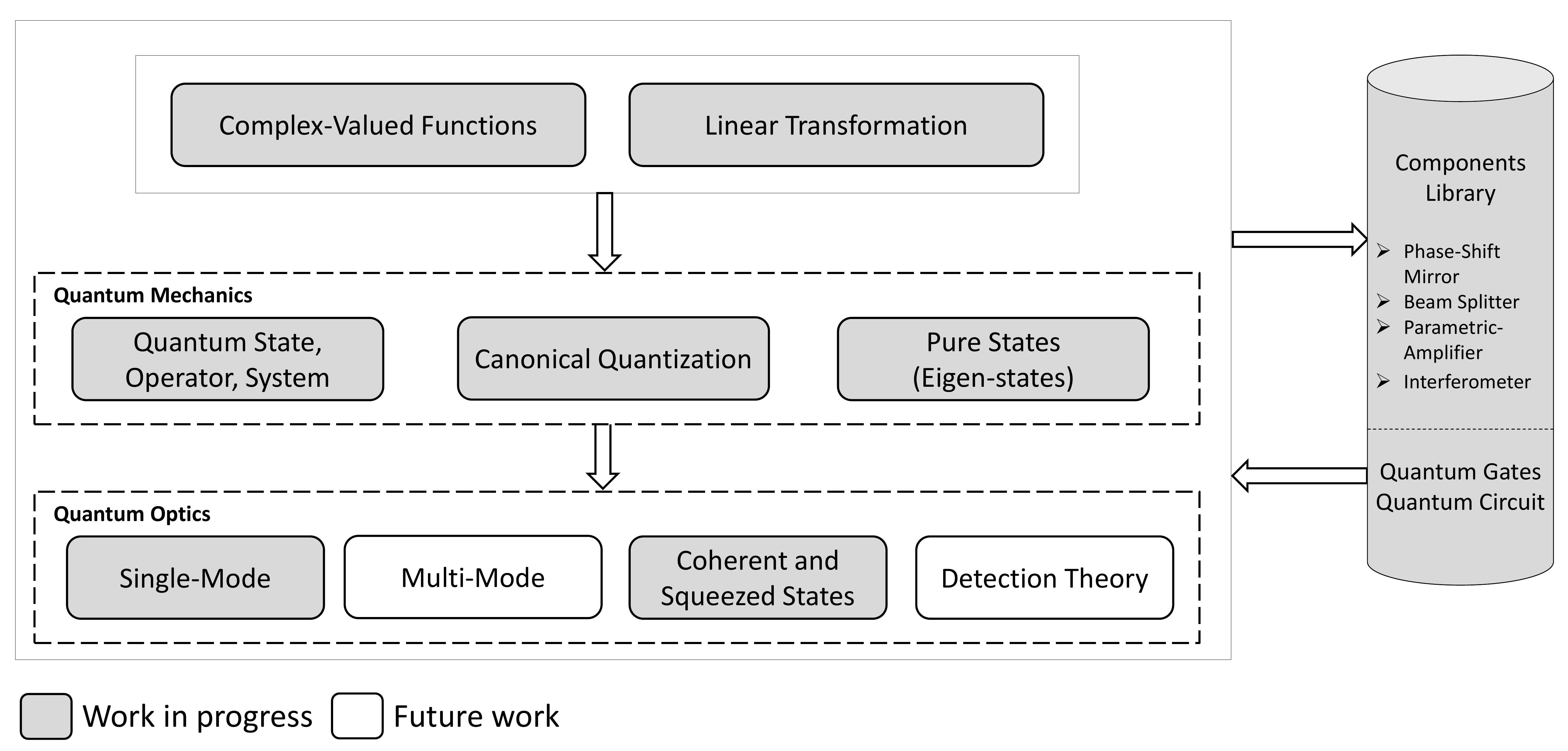}\vspace{-.3cm}
	\caption{Quantum Optical Systems Formalization Methodology}
	\label{fig:quantum}
\end{figure}

Figure \ref{fig:quantum} depicts the proposed formalization flow. It simply summaries the dependencies  among three essential theories in our work: 1) Linear algebra of complex-valued function, 2) Quantum mechanics and 3) Quantum optics.

As we summarized in the above section, each quantum notion requires some mathematical foundations, in particular, linear space of complex-valued function (i.e., infinite complex spaces), inner product, and transformation over complex function spaces, linearity and self-disjointness.
To the best of our knowledge, only four significant formalizations of linear algebra  exist in the literature so far:
two in HOL Light (\cite{Harrison_96} and \cite{sanaz}), one in PVS \cite{PVSlinag}, and one in Coq \cite{linalgCoq}.
The former three focus essentially on $n$-dimensional Euclidean and complex spaces,
whereas our work generalizes it to (possibly) infinite-dimension vector spaces of complex numbers
(more precisely, complex-valued-function spaces). More details about this formalization are  available in \cite{NFMalg}.

The foundational notions of  quantum mechanics and quantum optics can be developed using our formalization of linear algebra and are described in the next sub-sections. Optical devices, such as the
 parametric amplifier (more devices are shown in Figure \ref{fig:quantum}), can in turn be formalized based on these foundations.

\subsection{HOL Light Implementation}

In the previous section, we have summarized the key points of quantum aspects. Now, we present the corresponding appropriate formalism:
 \begin{itemize}
   \item Since complex-valued functions are not available, we start by defining a new type  $\mathtt{cfun:A\to}$\  \texttt{complex}, where \texttt{A}
   represents an arbitrary data type. This concept gives flexibility to our formalization since we can  provide general results that are valid for any system.
   \item Since we have a new type, we then have to define operations for  variables of this type. For a quantum state, there are two basic operations, i.e., addition and scalar-multiplication
       \end{itemize}
\begin{flushleft} \begin{mdframed}
       \begin{definition}[\hol{cfun} arithmetic]\vspace{1pt} \ \\
            \shol{ cfun\_add\ (v_1:cfun)\ (v_2:cfun)\ :cfun = \lambda x:A.\ v_1\ x + v_2\ x}\vspace{1pt}\\
           \shol{ cfun\_smul\ (a:\complex)\ (v:cfun)\ :cfun  = \lambda x:A.\ a * v\ x }
      \end{definition}
\end{mdframed} \end{flushleft}

\noindent Note that the addition operation is different from scalar addition. Now, we can build out of these operations a vector space for quantum states:
\begin{flushleft} \begin{mdframed}
 \begin{definition}[Complex-Valued Functions Space ]\vspace{1pt} \ \\
\shol{\vdashdef  is\_cfun\_subspace\ (spc:cfun \rightarrow bool) \Leftrightarrow}  \vspace{1pt}\\
 \shol{ \hspace{.8cm}\Forall{ x, y}  } \vspace{1pt}\\
 \shol{ \hspace{.8cm} x\ IN\ spc \wedge y\ IN\ spc \Rightarrow\  x+y\ IN\ spc \wedge (\forall\ a.\ a\ *\ x\ IN\ spc) \wedge cfun\_zero\ IN\ spc}
\end{definition}
%\vspace{-.5mm}
\end{mdframed} \end{flushleft}
\bigskip
  Accordingly, We define an inner product over the quantum states space as follows:

\begin{flushleft} \begin{mdframed}
 \begin{definition}[Inner Product Function ]\vspace{1pt} \ \\
\shol{\vdashdef  is\_inprod\ (inprod:cfun\to cfun\to complex) \Leftrightarrow}  \vspace{1pt}\\
 \shol{ \hspace{.8cm}\Forall{ x, y,z}  } \vspace{1pt}\\
 \shol{ \hspace{.8cm} cnj\ (inprod\ y\ x) = inprod\ x\ y\ \wedge}\vspace{1pt}\\
 \shol{ \hspace{.8cm} inprod\ (x+y)\ z = inprod\ x\ z + inprod\ y\ z \ \wedge}\vspace{1pt}\\
 \shol{ \hspace{.8cm} real\ (inprod\ x\ x) \wedge 0 \leq real\_of\_complex\ (inprod\ x\ x) \ \wedge}\vspace{1pt}\\
 \shol{ \hspace{.8cm}(inprod\ x\ x = Cx(0) \Rightarrow x = cfun\_zero)\ \wedge}\vspace{1pt}\\
 \shol{ \hspace{.8cm}\Forall {a}inprod\ x\ (a\ *\ y) = a * (inprod\ x\ y)}
\end{definition}
%\vspace{-.5mm}
\end{mdframed} \end{flushleft}

 \noindent where $\mathtt{cfun\_zero}$ is defined as $\lambda\ x.\ 0\ $ (i.e., it returns 0 irrespective of the input $x$). Note that we does not restrict ourselves to the integral function (as an inner product of quantum state space) and make it general. They above properties are enough to formalize quantum mechanics related notions.

 A quantum states space is then defined as follows:
\begin{flushleft} \begin{mdframed}
 \begin{definition}[Quantum State Space]\vspace{1pt} \ \\
\shol{is\_qspace\ ((vs,inprod):qspace)\ \Leftrightarrow\ is\_cfun\_subspace\ vs \wedge is\_inprod\ inprod}
\end{definition}
\end{mdframed} \end{flushleft}
\noindent Just like the inner product, we do not consider all properties of  quantum space since it is mathematically much more complicated than this. The definition caters for the properties that are required for theorems subject to prove.

The other essential concept of quantum mechanics is quantum operators, for which we define a  new type:  $\mathtt{cop:cfun\to cfun}$. Here, we give an example of the properties that a quantum operator attains, linearity:
\begin{flushleft} \begin{mdframed}
\begin{definition}[ Linear Transformation]\vspace{1pt} \ \\
\shol{is\_linear\_op \ op \Rightarrow \ \Forall{x,y,a}\ op \ (x + y) = op \ x + op \ y \ \wedge\ op\ (a * x) = a * (op \ x)}
\end{definition}
\end{mdframed} \end{flushleft}

Now, we formalize the single-mode field which utilizes all above presented formal definitions:
\begin{flushleft} \begin{mdframed}
\begin{definition}[Single-Mode Field]\vspace{1pt} \ \\
\label{def:sm}
\shol{is\_sm\ ((qs,cs,H),\omega:sm) \Leftrightarrow}\vspace{1pt} \\
\shol{is\_qsys\ (qs,cs,H) \wedge\  0 < \omega\ \wedge \ \Exists{q,p}cs = [q;p]\ \wedge\  H\ = Cx (\frac{\omega^2}2)\ *\ (q^2) + Cx (\frac 12)\ *\ (p^2)}
\end{definition}
\end{mdframed} \end{flushleft}
\noindent Besides the above mentioned definitions, we have formalized many other important definitions along with their corresponding properties related to quantum optics. For example,  the zero point energy theorem for a single mode field, which states that a field always contains energy even though there is no flux or charges.

In a nutshell, we believe that we have enough mathematical and physical foundations that allow us to formalize quantum devices and complete quantum systems.
Some of our formalization details can be found in \cite{NFMalg}, in addition to the formalization of  beam splitters which are commonly used in building quantum computers \cite{coherentcomputer}. Our future work is to extend our library to  some new  optical elements, e.g., Mach-Zehnder interferometer \cite{pereira_09}, photon detectors \cite{Hadfield_09}, and parametric amplifiers \cite{simpledevices} that are used  for building quantum gates and quantum networks.

\section {Application: Stability Analysis of Two-Mirror Fabry P\'{e}rot Resonator}
\label{sec_app}
  The use of optics yields smaller components, high-speed communication and huge information capacity. This provides the basis of miniaturized complex engineering systems including digital cameras,
 high-speed internet links, telescopes and satellites. Optoelectronic and laser
devices based on optical resonators \cite{Fund_of_Photonics} are fundamental building-blocks for new
generation, reliable, high-speed and low-power optical systems. Typically, optical resonators are used in lasers \cite{Siegman-Lasers}, optical bio-sensors \cite{bio-sensors_12}, refractometry \cite{CANCER-CELL-FP} and
reconfigurable wavelength division multiplexing-passive optical network (WDM-PON) systems \cite{WDM-2006}.

One of the most important design requirements of optical resonators is the stability which states that the beam of light remains within the optical resonator even after $N$ round-trips (essentially, $N$ can be infinite).
In fact, stability depends on the properties and arrangement of its components, e.g., curvature of mirrors or lenses, distance between the mirrors, and refractive index of the mirrors. Optical resonators are mostly modelled using the principles of
ray optics \cite{Fund_of_Photonics}.

In general, resonators differ by their geometry and components (interfaces and mirrors) used in their design.
Optical resonators are broadly classified as stable or unstable.   Stability analysis identifies geometric constraints of the optical components which ensure
that light remains inside the resonator. Both stable and unstable resonators have
diverse applications, e.g., stable resonators are used in the measurement of refractive index of cancer cells \cite{CANCER-CELL-FP}, whereas unstable resonators are used in the laser oscillators for high energy  applications \cite{Siegman-Lasers}.

\begin{figure}[h]
  \centering
{\includegraphics[width=12cm]{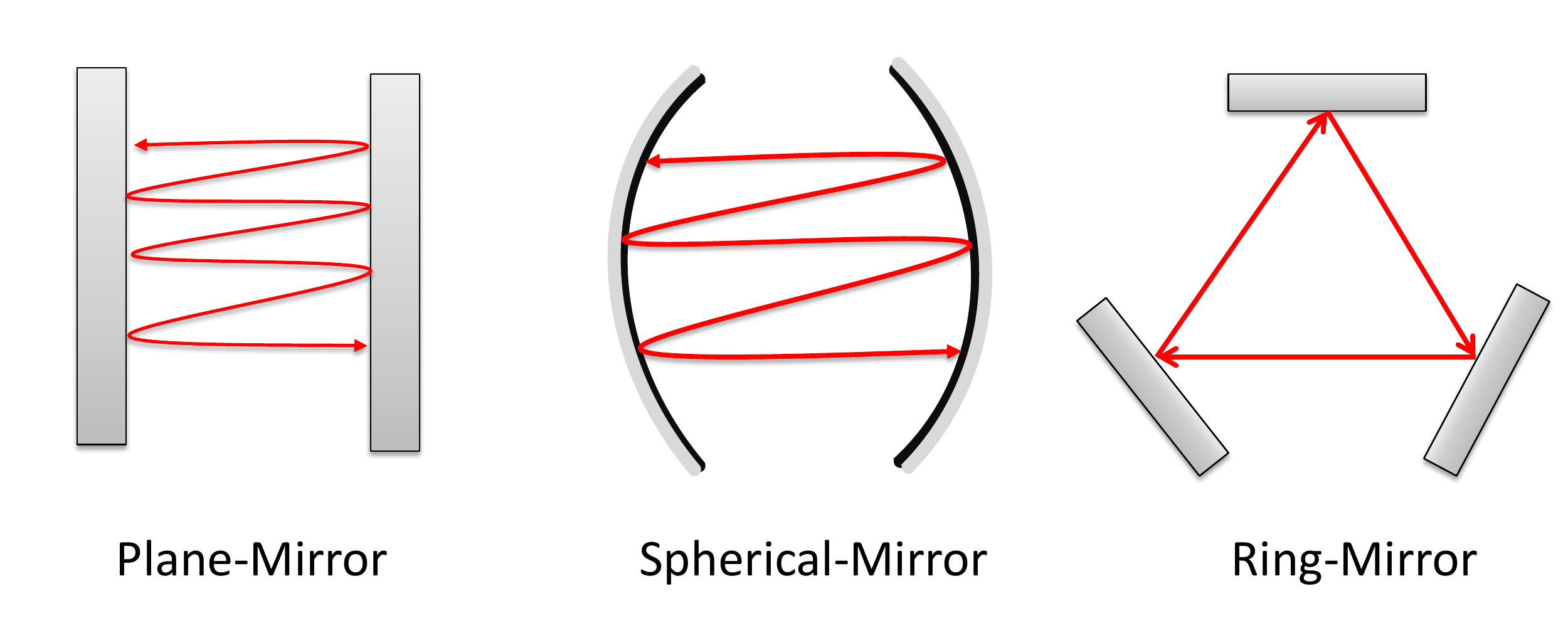}}
\caption{Optical Resonators}
\label{fig:gen-resonator} % labeling to refer it inside the text
\end{figure}

 The stability analysis of optical resonators involves the consideration of infinite rays, or, equivalently, of an infinite set of finite rays. Indeed, a resonator is a closed structure terminated by two reflected interfaces and a ray reflects back and forth between these interfaces.
For example, consider a simple plane-mirror resonator as shown in Figure \ref{fig:gen-resonator}: let $m_1$ be the first mirror, $m_2$ the second one, and $f$ the free space in between. Then the stability analysis involves the study of the ray as it goes through $f$, then reflects on $m_2$, then travels back through $f$, then reflects again on $m_1$, and starts over. So we have to consider the ray going through the ``infinite'' path $f,m_2,f,m_1,f,m_2,f,m_1,\dots$, or, using regular expressions notations, $(f,m_2,f,m_1)^*$. Our purpose, regarding stability, is to ensure that this infinite ray remains inside the cavity. This is equivalent to the fact that, for every $n$, the ray going through the path $(f,m_2,f,m_1)^n$ remains inside the cavity. This allows reducing the study of an infinite path to an infinite set of finite paths.

 In order to consider an infinite set of finite-path rays, we should thus consider an infinite set of optical systems. This has been naturally achieved by optics engineers by ``unfolding'' the resonator as many times as needed, depending on the considered ray. For instance, consider again the above example of a plane-mirror resonator: if we want to observe a ray going back and forth only once through the cavity, then we should consider the optical system made of $f,m_1,f,m_2$; however, if we want to study the behavior of rays which make two round-trips through the cavity, then we consider a \emph{new} optical system $f,m_1,f,m_2,f,m_1,f,m_2$; and similarly for more round-trips. This is the standard way in which the optics engineers handle resonators and chosen for our formalization, which we present now.

In our formalization, we want the user to provide only the minimum information so that HOL Light generates automatically the unfolded systems. Therefore, we do not define resonators as just optical systems but define a dedicated type for them: in their most general form, resonators are made of two reflecting interfaces and a list of components in between. We thus define the following type:

\vspace{5pt}
\begin{flushleft} \begin{mdframed}
 \begin{definition}[Optical Resonator] \label{D_int}\vspace{1pt} \ \\
\shol{define\_type\ ``resonator =  }\\
\shol{\hspace{1cm}:interface \times optical\_component\ list \times free\_space \times interface"}
\end{definition}
\end{mdframed} \end{flushleft}

Note that the additional free space in the type Definition \ref{D_int} is required because the type \shol{optical\_component} only contains one free space (the one before the interface, not the one after).

As usual, we introduce a predicate to ensure that a value of type \shol{resonator} indeed models a real resonator:

\vspace{5pt}
\begin{flushleft} \begin{mdframed}
  \begin{definition} [Valid Optical Resonator]
    \vspace{1pt} $\mathtt{}$\\
	\shol{\vdashdef\Forall {i_1\ cs\ fs\ i_2}}\\
	\shol{\hspace{.8cm}  is\_valid\_resonator\ ((i_1,cs,fs,i_2):resonator) \Leftrightarrow}\\
	\shol{\hspace{.8cm}is\_valid\_interface\ i_1\ \wedge\  ALL\ is\_valid\_optical\_component\ cs\ \wedge}\\
     \shol{\hspace{.8cm}is\_valid\_interface\ i_1\ \wedge\  ALL\ is\_valid\_optical\_component\ cs\ \wedge} \\
    \shol{\hspace{.8cm}is\_valid\_free\_space\ fs\  \wedge\  is\_valid\_interface\ i_1}
  \end{definition}
\end{mdframed} \end{flushleft}

We can now define formally the notion of stability.
 For an optical resonator to be stable, the distance of the ray from the optical axis
and its orientation should remain bounded whatever is the value of $N$. This is formalized
as follows:
 \begin{flushleft} \begin{mdframed}
 \begin{definition}[Resonator Stability] \label{def:stability}\vspace{1pt}\ \\
 \shol{\vdashdef  \Forall {res}}\\
 \shol{\hspace{.8cm} is\_stable\_resonator\ res \Leftrightarrow} \\
 \shol{\hspace{.8cm}  (\Forall r \Exists {y\ \theta} \Forall N\ is\_valid\_ray\_in\_system\ r\ (unfold\_resonator\ res\ N) \Rightarrow} \\
 \shol{\hspace{2.8cm}(let\ y_n,\theta_n = last\_single\_ray\ r\ in\
                             abs(y_n) \leq y \ \wedge\ abs( \theta_n)\ <\ \theta))}
  \end{definition}
\end{mdframed} \end{flushleft}

\noindent where, \shol{unfold\_resonator} accepts two parameters, i.e., a resonator (\shol{res}) and a number (\shol{N}) which specifies the
 number of round trips.

Formally proving that a  resonator satisfies the abstract condition of Definition \ref{def:stability} does not seem trivial at first.
However, if the determinant of a resonator matrix $M$ is $1$ (which is the case in practice),
optics engineers have known for a long time that having $-1<\frac{M_{11}+M_{22}}2<1$ is sufficient to ensure that the stability condition holds.
This can actually be proved by using Sylvester's Theorem  \cite{Sylvester_original}, which has already been formalized in  \cite{ADG_LNAI_UMAIR}.
Finally, we derive  the generalized stability theorem for any resonator as follows:
 \begin{flushleft} \begin{mdframed}
\begin{theorem}[Stability Theorem] \label{TH:stability}\vspace{1pt}\ \\
\shol{ \vdash \Forall {res}}\\
 \shol{\hspace{.3cm} is\_valid\_resonator\ res\ \wedge}\\
\shol{\hspace{.3cm} \Forall N let\ M = system\_composition\ (unfold\_resonator\ res\ 1)\ in}\\ %[0.3em]
\shol{\hspace{.3cm} det\ M = 1\ \wedge\ -1\ <\ \frac{M_{1,1} + M_{2,2}}{2}\ \wedge\ \frac{M_{1,1} + M_{2,2}}{2} < 1)  \Rightarrow is\_stable\_resonator\ res}
\end{theorem}
\end{mdframed} \end{flushleft}

\noindent where \shol{M_{i,j}} represents  the element at column \shol i and row \shol j of the matrix and \shol{det} represents determinate of
a matrix.
The formal verification of Theorem \ref{TH:stability} requires the definition of stability (Definition \ref{def:stability})  and  Sylvester's theorem \cite{ADG_LNAI_UMAIR}.
Note that our stability theorem is quite general and can be  used to verify the stability of almost all kinds of  optical resonators.

\begin{figure}[h]
  \centering
{\includegraphics[width=5cm]{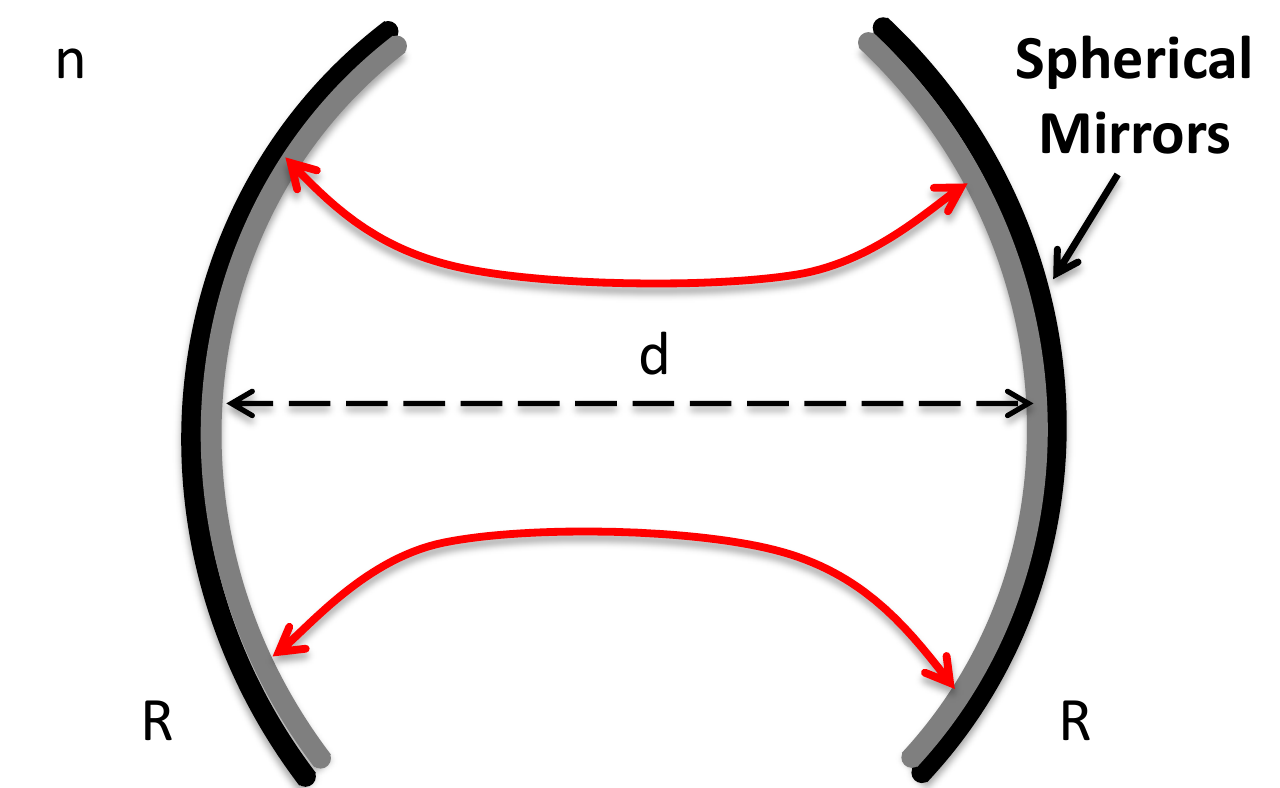}}
\caption{ Fabry P\'{e}rot Resonator}
\label{fig:fp-simple} % labeling to refer it inside the text
\end{figure}

As a direct application of the framework developed in this section, we present the stability analysis of the Fabry P\'{e}rot (FP) resonator with spherical mirrors as shown in Figure \ref{fig:fp-simple}. This architecture  is composed of two spherical mirrors with radius of curvature \shol R separated by a distance \shol d and refractive index \shol n.

We formally model this resonator as follows:
 \begin{flushleft} \begin{mdframed}
\begin{definition}[FP Resonator] \vspace{1pt}\  \\
\shol{\vdashdef \Forall{R\ d\ n}}\\
\shol{\hspace{.8cm} (fp\_resonator\ R\ d\ n\ :resonator) = (spherical\ R,[\ ],(n,d),spherical\ R)}
  \end{definition}
 \end{mdframed} \end{flushleft}
\noindent where \shol{[\ ]} represents an empty list of components because the given structure has no component between spherical interfaces but only a free space (n,d). Next, we verify that the FP resonator is indeed a valid resonator as follows:

 \begin{flushleft} \begin{mdframed}
\begin{theorem}[Valid FP resonator] \vspace{1pt}\ \\
\shol{\vdash \Forall{R\ d\ n} R \neq 0 \ \wedge\  0 \le d\  \wedge\ 0 <  n \Rightarrow is\_valid\_resonator\ (fp\_resonator\ R\ d\ n)}
  \end{theorem}
 \end{mdframed} \end{flushleft}
Finally, we formally verify the stability of the FP resonator as follows:.
 \begin{flushleft} \begin{mdframed}
\begin{theorem}[Stability of FP Resonator]\vspace{1pt}\ \\
\shol{\vdash \Forall{R\ d\ n}} \\
\shol{\hspace{.3cm} R \neq 0 \ \wedge\ 0 <  n\ \wedge\ 0 < \frac{d}{2} \ \wedge\ \frac{d}{2} < 2 \Rightarrow is\_stable\_resonator\ (fp\_resonator\ R\ d\ n)}
\end{theorem}
 \end{mdframed} \end{flushleft}

\noindent The first two assumptions just ensure the validity of the model description.
The two following ones provide the intended stability criteria.
The formal verification of the above theorem  requires Theorem \ref{TH:stability}  along with some fundamental properties of the matrices and arithmetic reasoning.

This completes our formalization of FP resonator which demonstrates the utilization of our ray optics formalization. We verified a generic result of
stability (Theorem \ref{TH:stability}) for any number of round trips and any considered ray of light within the resonator. Informally, this is achieved by tracing a
single ray for hundreds  of simulation runs which is of-course both time consuming and incomplete. On the other hand, formal verification of stability in a theorem prover provides explicit conditions (in the form of ranges of systems parameters, e.g., Theorem \ref{TH:stability}) under which the given resonator can be stable. This can reduce the problem of checking the stability to just the satisfaction of such conditions. We admit the fact that formalization of ray optics
requires a significant amount of time along with the expertise of both HOL and underlying physical concepts. We believe that such efforts paid off
when the time
required to analyze practical applications reduces to the fraction of the time to formalize required HOL theories. For example, analysis of
FP resonator requires around $150$ lines of HOL Light code and  $2$ man-hours by an expert user.

Apart from the above described  application,  we have developed a library of frequently used optical components,  such as thin lens, thick lens and dielectric plate (detailed description and formalization can be found in \cite{ADG_LNAI_UMAIR}). We showed the effectiveness of developed theories by the formal analysis of some more practical optical resonators like, Fabry P\'{e}rot resonator with Fiber-rod lens  and Z-shaped resonators \cite{umair_nfm,ADG_LNAI_UMAIR}. Moreover, we devised a generalized procedure for the formal stability analysis of optical resonators usable by physicists and optical engineers (details can be found in \cite{umair_spie}).

\section{Combining HOL Light and Mechanized Mathematical Systems}\label{sec_bridge}

The modelling and analysis of optical systems sometimes involves situations where
underlying mathematical equations have no  closed-form solution. Similarly, such an analysis involves the simplification of
complex mathematical expressions involving multivariate calculus. The first problem can be addressed by using well-known numerical techniques
which are readily integrated in computational tools such as MATLAB \cite{MATLAB}. On the other hand, the latter can be addressed using computer algebra  systems (e.g., Mathematica and Maple) which are
considered to be most efficient tools for computing symbolic solutions automatically. Both of these techniques have some known limitation of
incompleteness and soundness in case of numerical techniques and computer algebra systems, respectively.  In this work, our main idea is to leverage upon the  expressive nature and soundness of higher-order logic theorem proving as much as possible, but at the same time it cannot provide a stand-alone solution.
 In order to handle this situation,
 we propose to provide a bridge connecting theorem provers with symbolic and numerical techniques based tools.
 In this bridge, the given equation is first transferred to a Computer Algebra System (CAS) in order to be simplified symbolically. If it is successfully simplified, then we transfer it to the theorem prover (ideally by first certifying the simplification) in order to pursue the proof process. In case the equation cannot be simplified symbolically, we have no option but to switch to numerical approaches.  Note that all the simplifications are performed by CAS built-in functions and we do not export axioms and simplification rules of HOL Light.

 Linking theorem provers with CASs is an active research field, which can be broadly classified into two categories: either both the theorem prover and the CAS directly communicate with each other \cite{Harrisson,IsMa,Mathscheme} or one is built inside the other~\cite{theorema,kaliszyk_07}. Both categories so far cannot provide a rich repository of both axiomatic and algorithmic theories. However, our approach aims at developing a problem-solving environment ~\cite{connecting} based on the integration and interaction between  multiple Mechanized Mathematical Systems (MMSs).

\begin{figure}[h]
\begin{center}
\includegraphics[width=10cm,trim=80 500 50 100,clip=true]{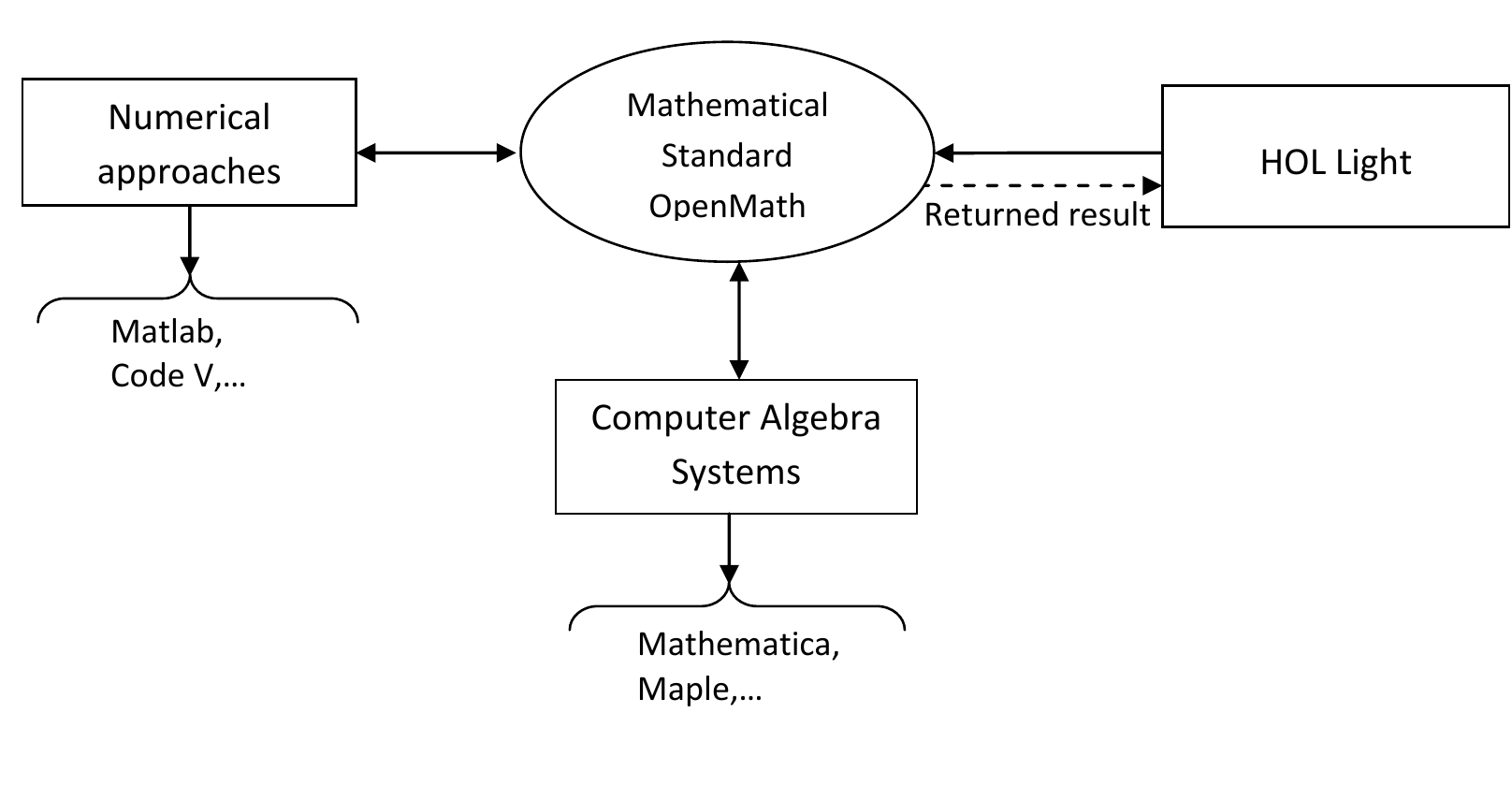}
\caption{Connecting Different MMSs using OpenMath}
\label{bloc_nous}
\end{center}
\end{figure}

Figure \ref{bloc_nous} explains our proposed architecture where different MMSs are connected.
The main goal of our work is to define a general approach to connect multiple MMSs together in a way to have access to their kernels. This approach provides us with a large number and a variety of different MMSs like theorem provers, CASs or numerical approaches with the intention of solving and reasoning over larger sets of problems. We propose to use OpenMath \cite{openmathweb} to connect different MMSs. OpenMath is a standard for representing mathematical objects with their semantics, allowing them to be exchanged between computer programs, stored in databases, or published on the worldwide web~\cite{openmathweb}.

\begin{figure}
\begin{center}
 \includegraphics[scale=.7,trim=50 350 50 100,clip=true]{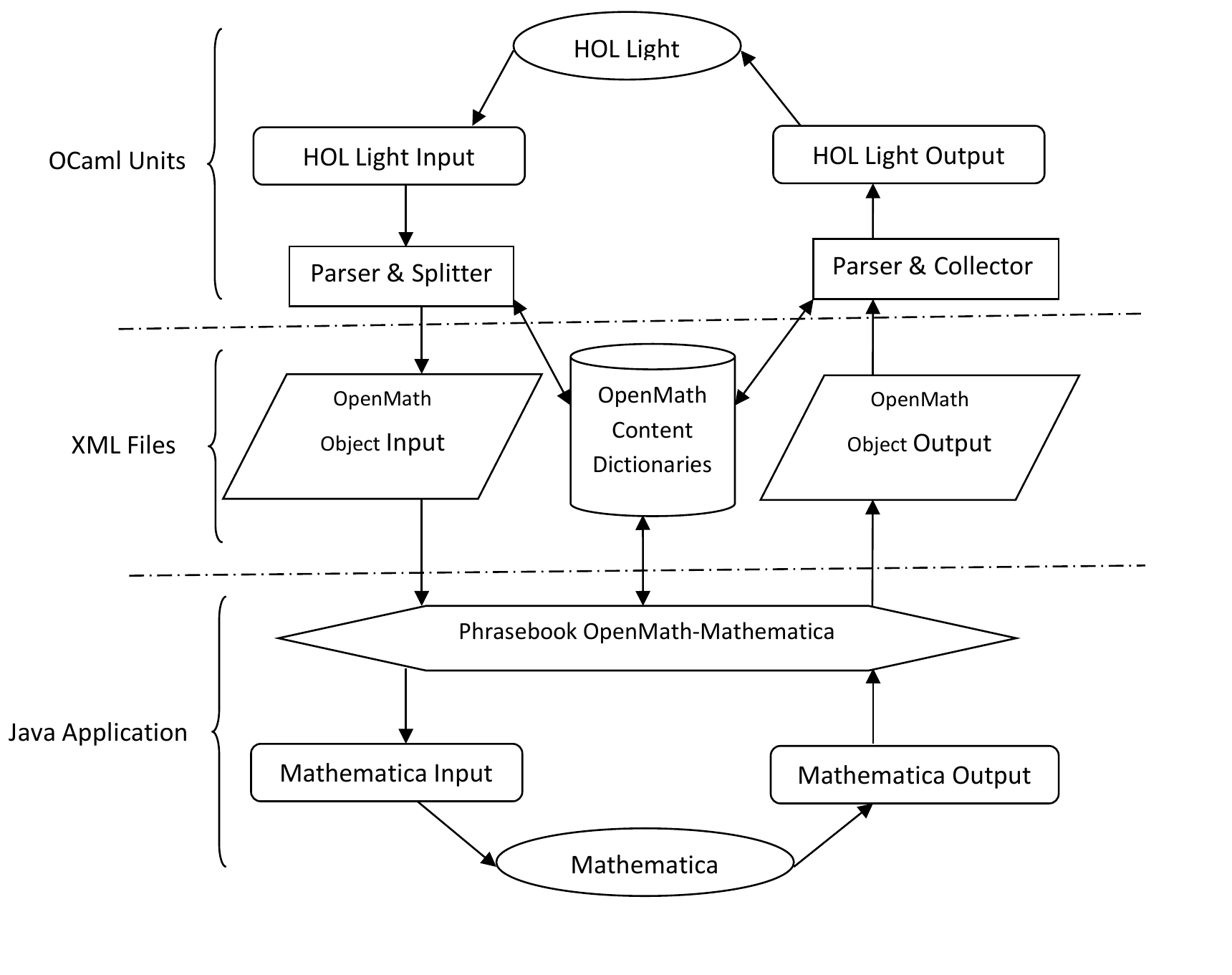}\vspace{-.3cm}
\caption{Flow of Connecting HOL Light to Mathematica}
\label{fr_nous}
\end{center}
\end{figure}

As a first step towards our  ultimate goal, we established a preliminary  link between HOL Light and Mathematica which supports both symbolic and numerical techniques.
Figure \ref{fr_nous}  depicts our methodology to connect HOL Light and Mathematica.
This is comprised of three modules: the OCaml units, the XML files which represent the OpenMath objects, and the Java application. It can be observed in Figure \ref{fr_nous} that the connection between each pair of modules is bidirectional. This makes the connection between HOL Light and Mathematica complete.
 If we have a HOL Light goal (or a mathematical expression) which needs to be simplified, we can call the main function \texttt{call\_mathematica}
 which accepts two arguments , i.e., the HOL Light expression as a string and the specific Mathematica function as a string.
 Then the HOL Light term is passed to the first module called ``Ocaml units''. It is composed of two components: ``Parser \& Splitter'' and ``Parser \& Collector''. At this stage,  we use the  ``Parser \& Splitter''  unit in order to  transform the HOL Light term into a corresponding OpenMath object. First, it parses and decomposes the HOL Light input statement into a list of operators and operands. Then, it maps each element of the list with the corresponding OpenMath  symbol  as decoded by means of the related Content Dictionaries (CDs)~\cite{cd}. Finally, it stores the description of the OpenMath object in an XML file.  A CD  describes the semantics of the mathematical object using a collection of symbols, their designations, their formal and informal descriptions, and rules which define the use of appropriate symbols in the correct order. This representation can be exchanged among several systems. The XML file is used as an input for the second module. In addition, the  ``Parser \& Splitter''  unit saves the tag of the specific Mathematica function in a text file in order to be passed to the Java application.

The second module is composed of two XML files that describe the mathematical objects: one file is the HOL Light input statement, the second file is the Mathematica output statement. At this stage we only have the OpenMath object description of the HOL Light input term. This file represents the input of the third module that contains the Java application which represents the {\it {phrasebook}} ~\cite{java}.

 The concept of the phrasebook between OpenMath and Mathematica (available from the Mathdox Web site \cite{code}) was introduced by Caprotti et al. ~\cite{java}. A phrasebook, by definition, is responsible for the translation of a mathematical expression expressed as an OpenMath object into its representation in the application ~\cite{mathopenm,connecting}, which in our case is Mathematica. This phrasebook is working under the latest version of  Mathematica (Mathematica 9.0 ~\cite{mathematica}). It provides a collection of encoding and decoding methods  between OpenMath and Mathematica based on the declaration of the corresponding CDs. Although it comprises of a large set of useful official and experimental CDs like: {\it {arith1}}, {\it {arith2}}, {\it {relation1}}, etc.~\cite{cd}, it does not include the recent ones such as linalg4 ~\cite{cd},  linalgeig1~\cite{linalgeig} and relation3 ~\cite{cd}. Before translating the XML description into an input Mathematica statement, we need to define the Mathematica calling function with the tag already specified by the user. Then,  we evaluate it through a connection with the MathLink~\cite{link} link which opens the Mathematica kernel. Once the result is computed, the Mathematica output statement is translated back to OpenMath; an XML file is generated and passed to the second module which is the XML files. The same process takes place in the reverse direction until we reach ``Parser \& Collector'' module which translates the OpenMath object into the specification of the HOL Light symbols in the relevant CDs.

Currently, we have been able to implement different functions such as simplifying transcendental expressions, computing eigenvalues, finding roots and
factorization of  complex polynomials. For example, we can simplify an expression as follows: \\
\begin{flushleft}
\texttt {\textbf{Input:}\\
\shol{\#call\_mathematica\ ``(x\ pow\ 2) + x + \&1 > (x\ pow\ 2) + x + \&2 "\ ``Simplify";;}\\
 \textbf{Output:}\\
 \shol{val\ it : thm = \ \vdash \neg(x\ pow\ 2 + x + \&1 > x\ pow\ 2 + x + \&2)} \\ }
\end{flushleft}
\vspace {0.25cm} \noindent Similarly, we can obtain the solution of an equation:
\begin{flushleft}
\texttt {\textbf{Input:}\\
\shol{\#call\_mathematica\ ``x\ pow\ 2 - \&1 = \&0"\ ``Solve";;}\\
\textbf{Output:}\\
\shol{val\ it : thm = \ \vdash x\ pow\ 2 - \&1 = \&0 \ \Longrightarrow \ x = - \&1 \  \vee \ x = \&1\ }\\}
\end{flushleft}
\bigskip

Note that these equations can  involve integrals, derivatives and transcendental functions. The current
implementation is still experimental and all the source codes and implementation details are available for download at \url{http://hvg.ece.concordia.ca/projects/optics/bridge.htm}.

\subsection*{Application: Boundary Condition of an Optical Interface}
To show the effectiveness of our approach and implementation to connect HOL Light and Mathematica, in this section we present the example of a boundary condition computing of an optical interface described with electromagnetic field.

\begin{figure}[!htb]
  \centering  % Requires \usepackage{graphicx}
  \includegraphics[width=2.2in]{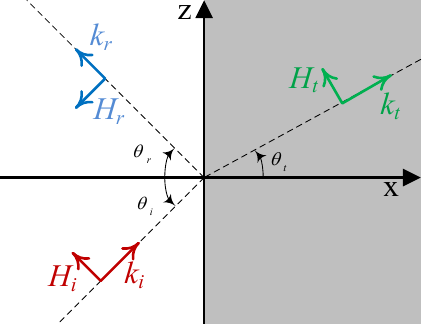}
  \caption{The system of Interface}\label{fig:CAS}
\end{figure}

As mentioned in Section \ref{sec_emf}, to ensure that a system can be expressed in our formal development of electromagnetic optics, it needs to satisfy a set of constraints. For example, in Figure \ref{fig:CAS}, the system of interface, which includes the interface \hol i and electromagnetic fields \hol {EMF^{(i)}}, \hol {EMF^{(r)}}, and \hol {EMF^{(t)}}, should satisfy the set of Constraints \ref{is_plane_wave}.\\

\vspace{0.3cm} Our objective is to validate one of the most important predicates of the set of Constraints \ref{is_plane_wave}: \hol {boundary\_conditions} (Definition \ref{bound_cond}).
	Noting that the interface in Figure \ref{fig:CAS} is the $yz$-plane, the objective is to validate that the electromagnetic fields  \hol {EMF^{(i)}}, \hol {EMF^{(r)}}, and \hol {EMF^{(t)}}  satisfy the boundary conditions. This can be formally described as follows:
\vspace{5pt}
\begin{flushleft} \begin{mdframed}
\begin{goal}[Boundary Conditions of the System of Interface in Figure \ref{fig:CAS}]   \vspace{1pt}
\label{goal_CAS_1}\ \\
\shol{\Forall {r\ t}}\\
\shol{let\ \theta_t = \arcsin (\frac{n_1}{n_2} \sin \theta_i)\ in\ let\  \theta_r = \theta_i \ in }\\
\shol{let\  r_a = (\frac{n_2 \cos \theta_i\  - n_1 \cos \theta_t}{n_2 \cos \theta_i\  +
       n_1 \cos \theta_t})a\ \ in\ \  \ let\  t_a = (\frac{2n_2 \cos \theta_i}{n_2 \cos \theta_i\  + n_1 \cos \theta_t})a\ in
       } \\\vspace{1pt}
\shol{ let\  k^{(i)} = k_0  n_1 [ \cos \theta_i;\ 0;\ \sin\theta_i]\ in  } \\
\shol{let\  k^{(r)} = k_0  n_1 [- \cos \theta_r;\ 0;\  \sin \theta_r]\ in} \\
\shol{ let\  k^{(t)} = k_0 n_2 [ \cos \theta_t;\ 0;\  \sin \theta_t]\ in} \\\vspace{1pt}
\shol{let\ E^{(i)} = [0;\ a e^{-j\ (k^{(i)}\qdot r - \omega t)};0]\ in} \\
\shol{let\ E^{(r)} = [0;\ r_a  e^{-j\ (k^{(r)}\qdot r - \omega t)};0]\ in} \\
\shol{let\ E^{(t)} = [0;\ t_a e^{-j\ (k^{(t)}\qdot r - \omega t)};0]\ in}\\ \vspace{1pt}
\shol{let\ H^{(i)} = [a (\frac{n_1}{eta_0}) \cos \theta_i \ e^{-j(k^{(i)}\qdot r - \omega t)};\ 0 ;\ -a (\frac{n_1}{eta_0}) \sin \theta_i\ e^{-j(k^{(i)}\qdot r - \omega t)}]\ in} \\
\shol{let\ H^{(r)} = [-r_a (\frac{n_1}{eta_0}) \cos \theta_r \ e^{-j(k^{(r)}\qdot r - \omega t)};\ 0 ;\ -r_a (\frac{n_1}{eta_0}) \sin \theta_r\ e^{-j(k^{(i)}\qdot r - \omega t)}]\ in} \\
\shol{let\ H^{(t)} = [t_a (\frac{n_2}{eta_0}) \cos \theta_t \ e^{-j(k^{(t)}\qdot r - \omega t)};\ 0 ;\ -t_a (\frac{n_2}{eta_0}) \sin \theta_t\ e^{-j(k^{(t)}\qdot r - \omega t)}]\ in} \\ \vspace{1pt}
\shol{let\ EMF^{(i)} = plane\_wave\ k^{(i)}\ \omega\ E^{(i)}\ H^{(i)}\ in  } \\
\shol{let\ EMF^{(r)} = plane\_wave\ k^{(r)}\ \omega\ E^{(r)}\ H^{(r)}\ in  } \\
\shol{let\ EMF^{(t)} = plane\_wave\ k^{(t)}\ \omega\ E^{(t)}\ H^{(t)}\ in  } \\
\hspace{0cm} \shol{\Forall{pt} pt\ IN\ (plane\_of\_interface\ i) \ \wedge\  (pt\$1 = 0)\ \Rightarrow} \\
\hspace{0.1cm} \shol{boundary\_conditions\ (EMF^{(i)}+EMF^{(r)})\ EMF^{(t)}\ (plane\_of\_interface\ i)\ r \ t }
\end{goal}
%\vspace{-.5mm}
\end{mdframed} \end{flushleft} \vspace{0.3cm}
\noindent The two variables \hol r and \hol t are indicating points in space and time, respectively.
The plane of interface is defined as \shol{\Forall{pt} pt\ IN\ (plane\_of\_interface\ i) \ \wedge\  (pt\$1 = 0)}, which is equivalent to the $yz$-plane or the plane described by ``\hol{x = 0}''. After simplifying the Goal \ref{goal_CAS_1}, we derive the expression shown in Goal \ref{goal_CAS2} which, for illustration purposes, we chose to send to Mathematica.

\vspace{0.3cm}
\begin{flushleft} \begin{mdframed}
\begin{goal}[The Expression Sent to Mathematica to be Simplified/Verified]   \vspace{1pt}
\label{goal_CAS2}\ \\

\shol{%vector
\ \ [1; 0; 0] \ ccross\ }\\
\shol{ \hspace{.2cm}%vector
([0\ ; a  e^{-j( k_0  n_1  z  \sin{\theta_i} -\omega  t)}; 0] \ +
}\\
\shol{\hspace{.2cm}%vector
 [0\ ; \frac{n_2  \cos{\theta_i} - n_1  \cos{(\arcsin (\frac{n_1}{n_2}
 \sin{\theta_i}))}}{n_2  \cos{\theta_i} + n_1  \cos {(\arcsin
(\frac{n_1}{n_2}  \sin{\theta_i})})}  a e^{-j( k_0  n_1  z
\sin{\theta_i} - \omega  t)};\ 0]) \  = } \\
 \shol{\hspace{.2cm} [1; 0; 0] \ ccross\ } \\
\shol{\hspace{.2cm} [0\ ; \frac{2  n_2  \cos{\theta_i}} {n_2  \cos{\theta_i} + n_1
\cos{(\arcsin (\frac{n_1}{n_2}  \sin{\theta_i})})}  a\ e^{-j( k_0  n_1
 z  \sin {(\arcsin (\frac{n_1}{n_2}  \sin{\theta_i})}) - \omega  t)};\
0]\ }\\
\shol{\hspace{.2cm} \vspace{.2cm} \bigwedge \vspace{.2cm}}\\
\shol{%vector
 \ \ [1; 0; 0]  \ ccross\ }\\
\shol{\hspace{.2cm}%vector
([a  \frac{n_1}{ \eta_0}  \cos {\theta_i}  e^{-j (k_0  n_1 z
\sin{\theta_i} - \omega  t)};0\ ; -a   \frac{n_1}{\eta_0}  \sin
{\theta_i}  e^{-j (k_0  n_1  z  \sin{\theta_i} - \omega  t)}] \ +}\\
\shol{ \hspace{.2cm}%vector
 [-\frac{n_2  \cos{\theta_i} - n_1  \cos {(\arcsin (\frac{n_1}{n_2}
\sin{\theta_i}))}}
           {n_2  \cos{\theta_i} + n_1 \cos {(\arcsin (\frac{n_1}{n_2}
\sin{\theta_i}))}}
           a \frac{n_1}{\eta_0} \cos {\theta_i} e^{-j(k_0  n_1  z
\sin{\theta_i} - \omega  t)};\ 0;}\\
\shol{ \hspace{.2cm}\ -\frac{n_2  \cos{\theta_i} - n_1  \cos {(\arcsin
(\frac{n_1}{n_2}  \sin{\theta_i}))}}
                   {n_2  \cos{\theta_i} + n_1 \cos {(\arcsin
(\frac{n_1}{n_2}  \sin{\theta_i}))}}
            a \frac{n_1}{\eta_0}  \sin {\theta_i} e^{-j(k_0  n_1  z
\sin{\theta_i} - \omega  t)}] \ =}\\
\vspace{0.2cm}
\shol{%vector
\ \ [1; 0; 0]  ccross\  }\\
\shol{\hspace{.2cm}%vector
[ \frac{2  n_2  \cos{\theta_i)}}{ n_2  \cos{\theta_i} + n_1  \cos{
(\arcsin (\frac{n_1}{n_2}\sin{\theta_i}))}}
 a \frac{n_2}{ \eta_0} \cos{ (\arcsin (\frac{n_1}{n_2}
\sin{\theta_i})} e^{-j(k_0  n_1  z  \sin{ (\arcsin (\frac{n_1}{n_2}
\sin{\theta_i}))} -\omega  t)}; \ 0;}\\
\shol{ \hspace{.2cm}\ - \frac{2  n_2  \cos{\theta_i}} {n_2
\cos{\theta_i} + n_1  \cos{(\arcsin (\frac{n_1}{n_2}
\sin{\theta_i}))}}
         a \frac{n_2}{\eta_0} \sin{(\arcsin (\frac{n_1}{n_2}
\sin{\theta_i}))} e^{-j(k_0  n_1  z  \sin{(\arcsin (\frac{n_1}{n_2}
\sin{\theta_i}))} - \omega  t)}])}

\end{goal}
%\vspace{-.5mm}
\end{mdframed} \end{flushleft}

\noindent Note that, in Goal \ref{goal_CAS2}, the predicate \hol{boundary\_conditions} is reduced to the cross product between the normal to the interface
(i.e., \shol{[1;0;0]}) and summation of electric fields or magnetic fields at the interface.
Now, we can validate the Goal \ref{goal_CAS2} by calling Mathematica which returns it as a theorem as follows: \\

\begin{flushleft}
\texttt {\textbf{Input:}}\\
\shol{\#call\_mathematica\ ``Statement \ of \ Goal \ \ref{goal_CAS2}" ``FullSimplify";;}\\
\texttt {\textbf{Output:}}\\
\shol{val\ it : thm = \ \vdash}\\
\shol{%vector
\ \ [1; 0; 0] \ ccross\ }\\
\shol{ \hspace{.2cm}%vector
([0\ ; a  e^{-j( k_0  n_1  z  \sin{\theta_i} -\omega  t)}; 0] \ +
 [0\ ; \frac{n_2  \cos{\theta_i} - n_1  \cos{(\arcsin (\frac{n_1}{n_2}
 \sin{\theta_i}))}}{n_2  \cos{\theta_i} + n_1  \cos {(\arcsin
(\frac{n_1}{n_2}  \sin{\theta_i})})}  a e^{-j( k_0  n_1  z
\sin{\theta_i} - \omega  t)};\ 0]) \  = } \\
 \shol{\hspace{.2cm} [1; 0; 0] \ ccross\ %} \\
%\shol{\hspace{.2cm}
[0\ ; \frac{2  n_2  \cos{\theta_i}} {n_2  \cos{\theta_i} + n_1
\cos{(\arcsin (\frac{n_1}{n_2}  \sin{\theta_i})})}  a\ e^{-j( k_0  n_1
 z  \sin {(\arcsin (\frac{n_1}{n_2}  \sin{\theta_i})}) - \omega  t)};\
0]\ }\\
\shol{\hspace{.2cm} \vspace{.2cm} \bigwedge \vspace{.2cm}}\\
\shol{%vector
 \ \ [1; 0; 0]  \ ccross\ }\\
\shol{\hspace{.2cm}%vector
([a  \frac{n_1}{ \eta_0}  \cos {\theta_i}  e^{-j (k_0  n_1 z
\sin{\theta_i} - \omega  t)};0\ ; -a   \frac{n_1}{\eta_0}  \sin
{\theta_i}  e^{-j (k_0  n_1  z  \sin{\theta_i} - \omega  t)}] \ +}\\
\shol{ \hspace{.2cm}%vector
 [-\frac{n_2  \cos{\theta_i} - n_1  \cos {(\arcsin (\frac{n_1}{n_2}
\sin{\theta_i}))}}
           {n_2  \cos{\theta_i} + n_1 \cos {(\arcsin (\frac{n_1}{n_2}
\sin{\theta_i}))}}
           a \frac{n_1}{\eta_0} \cos {\theta_i} e^{-j(k_0  n_1  z
\sin{\theta_i} - \omega  t)};\ 0;}\\
\shol{ \hspace{.2cm}\ -\frac{n_2  \cos{\theta_i} - n_1  \cos {(\arcsin
(\frac{n_1}{n_2}  \sin{\theta_i}))}}
                   {n_2  \cos{\theta_i} + n_1 \cos {(\arcsin
(\frac{n_1}{n_2}  \sin{\theta_i}))}}
            a \frac{n_1}{\eta_0}  \sin {\theta_i} e^{-j(k_0  n_1  z
\sin{\theta_i} - \omega  t)}] \ =}\\
\vspace{0.2cm}
\shol{%vector
\ \ [1; 0; 0]  ccross\  }\\
\shol{\hspace{.2cm}%vector
[ \frac{2  n_2  \cos{\theta_i)}}{ n_2  \cos{\theta_i} + n_1  \cos{
(\arcsin (\frac{n_1}{n_2}\sin{\theta_i}))}}
 a \frac{n_2}{ \eta_0} \cos{ (\arcsin (\frac{n_1}{n_2}
\sin{\theta_i})} e^{-j(k_0  n_1  z  \sin{ (\arcsin (\frac{n_1}{n_2}
\sin{\theta_i}))} -\omega  t)}; \ 0;}\\
\shol{ \hspace{.2cm}\ - \frac{2  n_2  \cos{\theta_i}} {n_2
\cos{\theta_i} + n_1  \cos{(\arcsin (\frac{n_1}{n_2}
\sin{\theta_i}))}}
         a \frac{n_2}{\eta_0} \sin{(\arcsin (\frac{n_1}{n_2}
\sin{\theta_i}))} e^{-j(k_0  n_1  z  \sin{(\arcsin (\frac{n_1}{n_2}
\sin{\theta_i}))} - \omega  t)}])}
\end{flushleft}

The total verification time for this proof on an Intel i7, 2.8GHz CPU with 32GB RAM
and Oracle Linux 6.3 OS was about 3 seconds, which includes the time to invoke Mathematica and get its feedback. This example illustrates the usefulness of the CAS link that greatly facilitates the interactive theorem proving process by automatically verifying some of the proof goals that would
have required hours of user guidance if verified by theorem proving alone.

\section{Engineering Prospects}\label{sec_CAS}

The  use of formal methods (particularly higher-order-logic theorem proving) in practical engineering system analysis is always very challenging. The main reason behind this is the unfamiliarity of formal methods in different engineering and physical sciences, e.g., control, mechanical and chemical engineering. To the best of our knowledge, formal methods have never been used in optical system analysis due to the above mentioned reasons. In this paper, we propose theorem proving as a complementary approach to the traditional optical system analysis techniques particularly, simulation and CASs. We believe that we have developed several important libraries for different optics theories to be used in the analysis of practical optical systems such as the design and verification of optical resonators (e.g., \cite{umair_nfm}).

After a detailed investigation of some popular optical system analysis tools \cite{rezonator,zemax,OPTICA,Synopsys}, we have
 identified some fundamental  requirements that should be considered in a formalization tool. The first requirement
 is the  analysis capabilities in all three domains, i.e., ray, electromagnetic and quantum optics.
 We believe that in our current framework, we have significantly advanced in the fulfilment of this requirement as described in
 Section \ref{sec_ray}, \ref{sec_emf} and \ref{sec_quantum}. The second requirement is  the
 availability of the most frequently used components, such as lenses, mirrors, resonators and quantum gates.
 Such a library can significantly reduce the analysis time which can ultimately assist in reducing the time-to-market of
new optical devices. In our current formalization, we have developed libraries of frequently used components in all three
domains (i.e., ray, electromagnetic and quantum optics) and this is a continuing process. The third requirement is
the facility to define some new component types which can be altogether a new type or a composition of existing components.
Given the expressiveness of higher-order logic and flexibility of our already defined data-types, this goal can easily be
achieved in our current framework. But this requires sufficient expertise of using our HOL Light developments, which is a drawback in the
context of targeted users of our formalization, i.e., physicists and optical engineers.

Most  optical system analysis tools \cite{rezonator,zemax,OPTICA,Synopsys} provide significant automation.
Due to the undecidable nature of HOL theorem proving, this is one of the most challenging requirements. We believe
that this particular requirement  depends on the nature of the properties of optical systems that
we need to analyze in a theorem prover. For example, proving stability can significantly be automated, but
proving the quantum efficiency of a resonator cannot be performed without user interaction.
Although during our formalization, we tried to automate as much tasks as we can by developing some automation tactics.
For example, in ray optics, \shol{VALID\_OPTICAL\_SYSTEM\_TAC} can automatically verify the validity of a given optical system.
Similarly, we have developed some  tactics in electromagnetic and quantum optics formalization and details can be
found in our source codes \cite{doform_home}.

Another important requirement is to perform computational analysis in case
     of having no closed-form solutions
        and symbolic analysis in case of evaluating some complicated mathematical expressions
        involving partial differential equations, eigenvalues and vector integrals.
 Nowadays, it is a common practice to translate optical system analysis problems from one tool to another (e.g., \cite{matlab_codev,matlab_zemax}).
 Providing such facilities in a formalization tool is not an easy task, however, we propose a preliminary framework to
 connect different formal and informal tools as described in Section \ref{sec_bridge}.
 Finally,  an essential requirement is to build  a graphical user interface (GUI) to provide an easy interaction between the user and the HOL Light theorem
 prover. Such a GUI can be built on arbitrary choices, i.e., text based or using computer aided design (CAD) tools. In each case, the main task is to properly translate the given information (text or a diagram) to match our HOL Light definitions. Recently, we have started focusing in this direction with a goal of building a tool for the formal stability
 analysis of optical resonators. The main intention is to assess the feasibility and use it as a base to build a comprehensive GUI to help researchers and practitioners involved in  optical system analysis.

 \section{Discussion and Conclusion} \label{sec_disc}
 In this paper, we presented a framework to enhance the  verification of optical systems, by introducing the formal analysis of optical system using higher-order-logic theorem proving as a complementary approach to traditional techniques, namely simulation based approaches and CASs. In this regard, we explained how to have both the system specification (in terms of properties) and its implementation (in terms of a physical model based on classical theories of optics or quantum optics) formalized. We can then verify, in HOL Light, that the implementation implies all the properties extracted from the specification.
Our framework certifies and clarifies the exact logical steps in which the physical assumptions and idealizations are pivotal inside each theory.
Invariably, such details are implicit or get lost in traditional approaches: one makes assumptions and idealizations exactly to forget about details and to obtain simpler mathematics to work on; using a computer even only to book-keep such details can add significant, reliable and readily usable knowledge to the whole theoretical framework. In our approach, due to the nature of formal reasoning, not only all the details of analysis are explicitly addressed, it guarantees that no assumption can be contradicted if a design evolves or is included as part of a hybrid system.

In practice, the verification of an optical system might require some modelling involving various theories, in a hierarchical  manner; for example, an early analysis and feasibility study is done using ray optics, then an advanced analysis concerning the properties of light wave is conducted by wave and electromagnetic optics. Finally, if the system involves light sources within its configuration then quantum optics is used.
Therefore, in order to make this task easier for verification engineers, we propose to wrap all the above formalizations in a uniform framework and to develop a mechanism to apply different theories on different aspects of a given system.
In order to achieve such common front-end, we developed a library of frequently-used optical components which are basic blocks of optical system models. For instance, in ray, wave and even electromagnetic optics, thin lenses, thick lenses and mirrors are such frequently used components. Some more complex systems can also be considered, e.g., various kinds of cavities, or resonators. In quantum optics, beam splitters are an example of such a basic component. Since quantum optics also has lots of other applications, e.g., in quantum computers, various devices like  quantum gates (which are to quantum computers what usual gates are to usual computers) would also be worth adding to such a library. In all cases, since such components are the basic blocks of optical systems, this library will help to formalize new optical systems. This is utterly important for the practical usability of the library.

Finally, considering the mathematical complexity of optical model analysis, we proposed to develop a bridge between HOL Light and Mathematica in order to increase our problem sets. The main advantage of our approach is that not only we are connecting the formal part of our framework to the one of the best available CASs in Optics, but given the popularity of OpenMath standards, we are hoping to have our bridge extended to many different engines, e.g., MuPAD, a symbolic toolbox of Matlab, which has already developed an OpenMath toolbox \cite{mupad}. However, relying on the results of CASs reduces the degree of accuracy in our approach and the challenge remains to maintain consistency of formalization in HOL Light while using CASs.

Note that the reported work is a part of our ongoing research project \cite{doform_home} of optical system verification. We believe that our expandable research
efforts will result in  enriched HOL Light libraries of optics and the development of some new proof technologies.

  \bibliographystyle{plain}
  \bibliography{all_new}

\end{document}